\documentclass[preprint,aps,showpacs,showkeys,amsmath,amssymb]{revtex4}

\usepackage{graphicx}
\usepackage{dcolumn}
\usepackage{bm}
\def\comment#1{}
\newcommand{\be}{\begin{equation}}\newcommand{\ee}{\end{equation}}
\newcommand{\bea}{\begin{eqnarray}}\newcommand{\eea}{\end{eqnarray}}
\newcommand{\brr}{\begin{array}}\newcommand{\err}{\end{array}}
\newcommand{\bit}{\begin{itemize}}\newcommand{\eit}{\end{itemize}}
\newcommand{\ben}{\begin{enumerate}}\newcommand{\een}{\end{enumerate}}

\renewcommand{\baselinestretch}{1.1}

\def\1{{_{1}}}\def\2{{_{2}}}
\def\d{\textrm{d}}

\begin{document}

\title{Perturbation Expansion for Option Pricing with Stochastic Volatility}%

\author{Petr Jizba}
\email{jizba@physik.fu-berlin.de}\altaffiliation[\\$\,$On leave
from~]{FNSPE, Czech Technical University, B\v{r}ehov\'{a} 7, 115 19
Praha 1, Czech Republic}
\author{Hagen Kleinert}%
 \email{kleinert@physik.fu-berlin.de}
\affiliation{%
ITP, Freie Universit\"{a}t Berlin, Arnimallee 14
D-14195 Berlin, Germany
}%

\author{Patrick Haener}
\email{patrick.haener@uk.nomura.com}
\affiliation{
Nomura International, Nomura House, 1 St Martin's-le-Grand, London, EC1A NP, UK
}%

\date{\today\\[5mm]}

\begin{abstract}
\vspace{3mm} We fit the volatility fluctuations of the S\&P 500
index well by a Chi distribution, and the distribution of
log-returns by a corresponding superposition of Gaussian
distributions. The Fourier transform of this is, remarkably, of the
Tsallis type. An option pricing formula is derived from the same
superposition of Black-Scholes expressions. An explicit analytic
formula is deduced from a perturbation expansion around a
Black-Scholes formula with the mean volatility. The expansion has
two parts. The first takes into account the non-Gaussian character
of the stock-fluctuations and is organized by powers of the excess
kurtosis, the second is contract based, and is organized by the
moments of moneyness of the option. With this expansion we show that
for the Dow Jones Euro Stoxx $50$ option data, a $\Delta$-hedging
strategy is close to being optimal.
\end{abstract}


\maketitle
%

\section{Introduction}

The purpose of this paper is to develop analytic expressions for
option-pricing of markets with fluctuating volatilities of a given
distribution. There are several good reasons for considering the
statistical properties of volatilities. For instance, a number of
important models of price changes include explicitly their
time-dependence, for example the Heston model~\cite{Heston1}, or the
famous ARCH~\cite{Engle82}, GARCH~\cite{Bollerslev86}, and
multiscale GARCH~\cite{Z-L03} models. Changes in the daily
volatility are qualitatively well explained by models relating
volatility to the amount of information arriving in the market at a
given time~\cite{Jacquier}. There is also considerable practical
interest in volatility distributions since they provide traders with
an essential quantitative information on the riskiness of an
asset~\cite{B-P,Bouchaud94}. As such it provides a key input in
portfolio construction.

\begin{figure}[h]
\hspace{-0.4cm}\centerline{\resizebox{17cm}{5cm}{\includegraphics{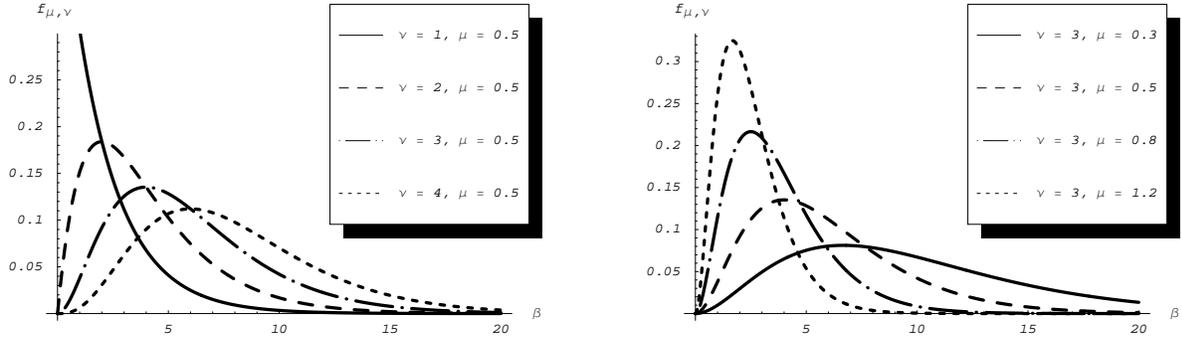}}}
\vspace{-0.6cm}\caption{\label{G.1} Gamma distribution for various
values of $\mu$ and $\nu$, with average $\bar v= \nu /\mu= \beta
$,\, variance $\overline{(v-\bar v)^2}= \nu /\mu^2$, skewness
$\overline{(v-\bar v)^3}= 2/ \sqrt{\nu} $, and excess kurtosis
$\overline{(v-\bar v)^4}/\overline{(v-\bar v)^2}^2-3=6/ \nu $. }
\vspace{0.2cm} \hrule
\end{figure}

 Unlike returns which are correlated only on very short time
scales~\cite{Fama70} of a few minutes and can roughly be
approximated by Markovian process, the volatility changes exhibit
memory with time correlations up to many
years~\cite{Ding93,Dacorogna93,Liu97,TIMX}. In Ref.~\cite{Liu97} it
was shown that the Standard \& Poor 500 (hereafter S\&P 500)
volatility data can be fitted quite well by a log-normal
distribution. This result appears to be at odds with the fact that
the log-normal shape of the distribution typically signalizes a
multiplicative nature~\cite{Bunde96} of an underlying stochastic
process. This is rather surprising in view of efficient market
theories~\cite{Fama70} which assume that the price changes are
caused by incoming new information about an asset. Such
information-induced price changes are additive and should not give
rise to multiplicative process. We cure this contradiction by
observing that the same volatility data can be fitted equally well
by a Chi distribution~\cite{feller,wik}. In Appendix~A we show that
corresponding sample paths follow the additive rather than
multiplicative It\={o} stochastic process. With the help of
It\={o}'s lemma one may show (cf. Appendix~A) that the variance
$v(t) = \sigma^2(t)$ follows the It\={o}'s stochastic equation
%
%
\begin{eqnarray}\nonumber \\[-4mm]
\d v(t) \ = \ \gamma(t)[\nu(t) - \mu(t) v(t) - a(v(t), \mu(t),
\nu(t))] \d t \ + \ \sqrt{2 \gamma(t) v(t)} \; \! \d W(t)\, .
\label{In.2}
\\[-3mm] \nonumber\end{eqnarray}
where $W(t)$ is a Wiener process. Here $\gamma(t), \mu(t)$ and $\nu(t)$
are arbitrary non-singular
positive real functions on $\mathbb{{R}}^+$. Function $a(\ldots)$ is
non-singular in all its arguments and it tends to zero at large $t$'s.
From the corresponding Fokker-Planck equation one may determine
the time-dependent distribution  of $v$ which reads
 \begin{eqnarray}\nonumber \\[-3mm]
f_{\mu(t),\nu(t)}(v) ~= ~ \frac{1}{\Gamma(\nu(t))} ~[\mu(t)]^{\nu(t)} v^{\nu(t)
-1} e^{-\mu(t) v}\, , \;\;\;\;\;\;\mbox{with}\;\;\;\;\; \int_0^{\infty} \d v \
f_{\mu,\nu}(v)  ~= ~1.\label{Gammad}
\\[-3mm] \nonumber\end{eqnarray}
The distribution $f_{\mu,\nu}(v)$ is the normalized Gamma
probability density~\cite{feller}, whose profile is shown in
Fig.~\ref{G.1}. It has an average  $\bar v= \nu /\mu$,\, a variance
$\overline{(v-\bar v)^2}= \nu /\mu^2$, a skewness $\overline{(v-\bar
v)^3}= 2/ \sqrt{ \nu } $, and an excess kurtosis $\overline{(v-\bar
v)^4}/\overline{(v-\bar v)^2}^2-3=6/ \nu $ (see, e.g.,
Refs.\cite{feller,HK}). The Gamma distribution (\ref{Gammad}) will
play a key role in the following reasonings. The Chi distribution
$\rho(\sigma,t)$ is related with the Gamma distribution through the
relation
 \begin{eqnarray}\nonumber \\[-3mm]
\rho(\sigma,t) \ = \ 2 \sigma f_{\mu(t),\nu(t)}(\sigma^2)\, .
\label{Chi}
\\[-3mm] \nonumber\end{eqnarray}
Often is the functional form (\ref{Chi}) itself called a Gamma distribution.
To avoid potential ambiguities,
we shall confine in the following to the name
Chi distribution.  The derivation of $\rho(\sigma,t)$
from the underlying additive process, rather than a multiplicative
process, shows that the Chi distribution is well compatible with efficient markets.

 The Gamma distributions for fluctuations of $v$ allow us to
generate an entirely new class of option pricing formulas. For this
we use the well-known fact of non-equilibrium statistical
physics~\cite{kubo},  that the density matrix of a system with
fluctuating temperature can be written as a density matrix for the
system with fixed temperature averaged with respect to a temperature
distribution function. Path integrals conveniently facilitate this
task~\cite{HK}. With the help of the so-called Schwinger trick we
show that if the distribution of the inverse-temperatures of the
log-returns is of the Gamma type, the distribution in momentum space
is of the Tsallis type. To put this observation into a relevant
context, we recall that Tsallis distributions in momentum space
enjoy a key role in statistical physics as being optimal in an
information theoretical sense: given  prior information only on  the
covariance matrix and a so-called escort parameter, they contain the
least possible assumptions, i.e., they are the most likely unbiased
representation of the provided data. Some background material on
this subject is reviewed in Appendix~C. The observed connection with
Tsallis distribution in momentum space can be fruitfully used to
address the issue of the time-compounded density function for stock
fluctuations with Chi distributed volatility. The Markovian
property of the market prices then simply translates to product of
distributions in momentum space. At the level of the distributions
of log-returns, this automatically gives rise to the correct
Chapman-Kolmogorov equation for Markovian processes. The
time-compounded distribution function constructed by convolution of
integrals represents the desired measure of stock fluctuations at
given time $t$.

 It should be emphasized that since the distribution of log-returns
is related to the Tsallis distribution in momentum space by a
Fourier transform, the log-return data do not inherit the heavy
tails of the Tsallis distribution in momentum space. In this respect
our approach is quite different from data analysis based on a
Tsallis distribution of log-returns, such as the so-called
`non-extensive thermostatic" \cite{Borland}. The formal advantage of
our model lies in its optimal use of the mathematical machinery of
the Black-Scholes theory. In particular, one still has the option
pricing formula that is linear in the spot probability of the
strike-price payment.
Put and call options still obey the put-call parity relation, and
the $\Delta$-hedge still coincides with the spot probability that
multiplies the asset price in the Black-Scholes formula. On the
other hand, the desired features such as peaked middles of financial
asset fluctuations and semi-fat tails are present for sufficiently
long times.
The latter ensures that
ensuing option prices can differ noticeably from the
Black-Scholes curves for rather long expiration dates (days or
even moths) despite the validity of central limit theorem (CLT). One
may thus expect  our formulation to be
useful, e.g., for short or mid-term maturity options (``mesoscopic"
time lags).

 The paper is organized as follows. In Section~\ref{SEc2a} we fit the
high-frequency data set of the volatility fluctuations in the S\&P
500 index from 1985 to 2007 to a Chi distribution. We also show
that this fit holds well for different time averaging windows. In
Section~\ref{SEc3} we represent the distribution of log-returns as a
superposition of Gaussian distributions with the above volatility
behavior and find that the associated distribution in momentum space
is of the  Tsallis type. By assuming further that the compounded
stock fluctuations over longer times are Markovian we obtain in
Section~\ref{opf} the corresponding natural martingale measure. With
it we derive an option-pricing formula as a superposition of
Black-Scholes expressions. The departure from Black-Scholes results
is expressed as a sum of two qualitatively different expansions:
expansion in powers of the excess curtosis and expansion in momenta
of moneyness. The former represents the expansion in market
characteristics while the latter is an expansion based on
contract characteristics. In Section~\ref{SEc8} we determine the
crossover time below which our option prices differs from
Black-Scholes. Comparisons with observed log-returns are presented
in Section~\ref{SEc6b}. There we show that for the Dow Jones Euro
Stoxx 50 data the amount of the residual risk in the $\Delta$-hedge
portfolio is very small and that the crossover time is roughly $7-8$
months. Section~\ref{SEc7}, finally, is devoted to conclusions. For
reader's convenience we also include three appendices. In Appendix~A
we solve and discuss the Fokker-Planck equations that are associated
with stochastic equations for the volatility and variance presented
in Introduction. In Appendix~B we deal with some mathematical
manipulations needed in Section~\ref{opf}. In Appendix~C we present
some basics for R\'enyi and THC statistics to better understand some
remarks in the main body of the paper.

\section{Empirical  motivation} \label{SEc2a}

 Our work is motivated by data sets on volatility distributions
extracted from the S\&P 500 stock market index. We briefly remind
the reader the procedure for obtaining these numbers. A detailed
description can be found in Ref.\cite{Liu97} and citations therein.

It is well known that the autocorrelation function of stock market
returns decays exponentially with a short characteristic time --
typically few trading minutes (e.g., $\approx 4$ min. for S\&P 500
index). Hence the log-returns $R$ are basically uncorrelated random
variables, nevertheless, they are not independent since higher-order
correlations reveal a richer structure. In fact, empirical analysis
of financial data confirms~\cite{Liu97} that the autocorrelation
function of non-linear functions of $R$, such as $|R|$ or $R^2$,
has much longer decorrelation time (memory) spanning up to several
years (few months for S\&P 500 index). These observations imply that
a realistic model for the return-generating processes should
account for a non-linear dependence in the returns. The latter can
be most naturally achieved via some additional memory-bearing
stochastic process. Most simply, this subsidiary process involves
directly the volatility.

One may bypass a construction of the volatility model by trying to
define a ``judicious" volatility estimator directly from financial
time series. Difficulty resides, however, in the fact that although
the volatility is supposed to be a measure of the magnitude of
market fluctuations, it is not immediately clear how such a measure
should be quantified. Among many definitions present in the
literature~\cite{pasquini98} we focuss here on the estimator
proposed in Ref.\cite{Liu97}. There one defines the volatility
$\sigma_T(t_{\eta})$, at a given time $t_{\eta}$ as the arithmetic
average of the absolute value of the log-returns
 \begin{eqnarray}\nonumber \\[-3mm]
R(t_n) \ \equiv \ \ln \frac{S(t_n + \Delta t )}{S(t_n)}\ \cong \
\frac{S(t_n + \Delta t ) - S(t_n)}{S(t_n)}  \, ,
\\[-3mm] \nonumber\end{eqnarray}
over some time window of the length $T \equiv n \Delta t$ ($n \in
\mathbb{N}^+$, $\Delta t$ is the sampling interval), i.e.,
 \begin{eqnarray}\nonumber \\[-3mm]
\sigma_T(t_{\eta}) \ \equiv \ \frac{1}{n} \sum_{m = \eta}^{n-1 +
\eta} |R(t_m)|\, ,\;\;\;\;\; (t_{\eta} = \eta \Delta t, \; t_m =
m\Delta t ) \, . \label{II.5}
\\[-3mm] \nonumber\end{eqnarray}
Because (\ref{II.5}) is basically a forward-time mean value of
absolute returns, it may indeed serve as a good measure of market
fluctuations.

With this $\sigma_T(t)$  we can construct the corresponding
volatility probability density function according to a relative
frequency prescription:
 \begin{eqnarray}\nonumber \\[-3mm]
\rho_T(\sigma) \Delta \sigma \ = \ \frac{\mbox{\# $\sigma_T(t) \in
[\sigma, \sigma + \Delta \sigma]$}}{n}\, .
\\[-3mm] \nonumber\end{eqnarray}
In the limit of the very large time window $T$, one might expect
$\rho_T(\sigma)$ to approach a Gaussian distribution, since the CLT
holds also for correlated time series~\cite{Beran94}, although with
an often  slower convergence than for independent
processes~\cite{Potters96}.

\begin{figure}[h]
\hspace{-0.4cm}\centerline{\resizebox{16cm}{6.8cm}{\includegraphics{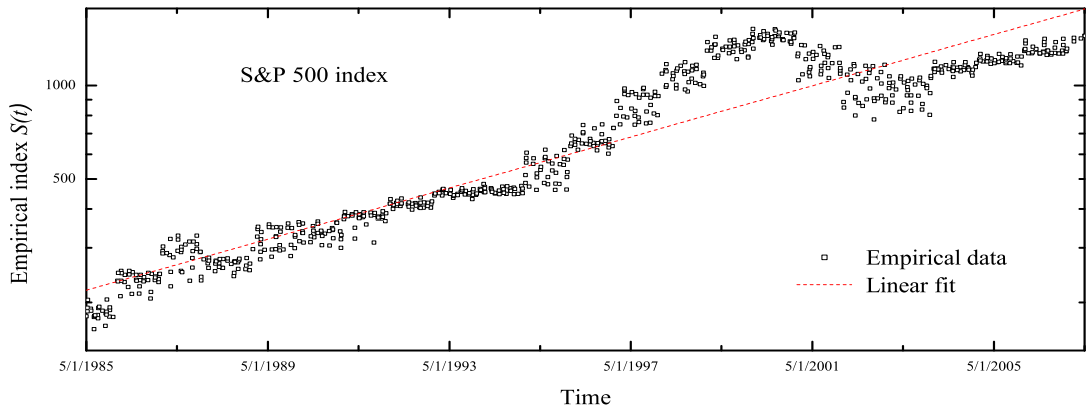}}}
\vspace{-0.9cm}\caption{\label{SP500index} Logarithmic plot of the
S\&P 500 index $S(t)$ over $22$ years (5 January 1985 - 5 January
2007) with sampling intervals $\Delta t = 5$ min. The linear fit
shows the typical exponential growth at an annual rate of $\approx
15\%$. Only end-of-day prices are shown.} \vspace{0.2cm} \hrule
\end{figure}

\begin{figure}[h]
\hspace{-0.4cm}\centerline{\resizebox{16.5cm}{7.4cm}{\includegraphics{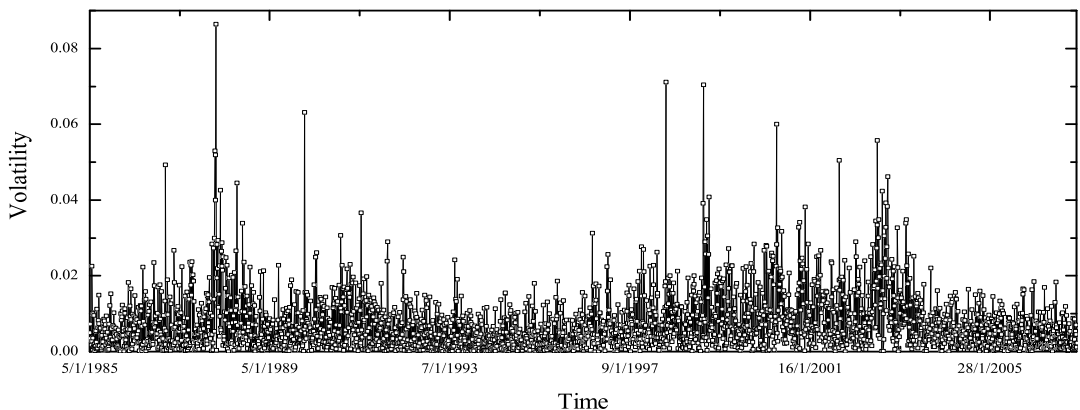}}}
\vspace{-0.9cm}\caption{\label{SP500vol-notnorm} Volatility
$\sigma_T(t)$ of S\&P 500 index over $22$ years (5 January 1985 - 5
January 2007) with sampling intervals $\Delta t = 5~$min. and time
window $T=300~$min., i.e., roughly one trading day. Only $3000$
volatility values are plotted.} \vspace{0.2cm} \hrule
\end{figure}

\begin{figure}[h]
\hspace{-0.4cm}\centerline{\resizebox{16cm}{7cm}{\includegraphics{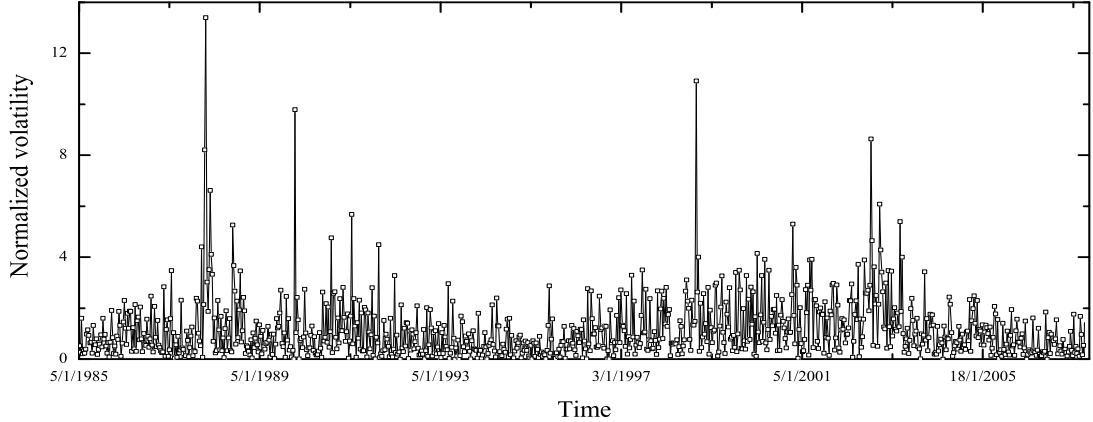}}}
\vspace{-0.6cm}\caption{\label{SP500vol} Normalized volatility
$\sigma_{T\rm{nor}}(t)$ of S\&P 500 index over $22$ years (5 January
1985 - 5 January 2007) with sampling interval $\Delta t = 5~$min.,
$N_{\rm{td}} = 5550$, and time window $T=300~$min., i.e., roughly
one trading day. Only $1000$ volatility values are plotted.}
\vspace{0.2cm} \hrule
\end{figure}

\begin{figure}[h]
\hspace{-0.4cm}\centerline{\resizebox{16cm}{7cm}{\includegraphics{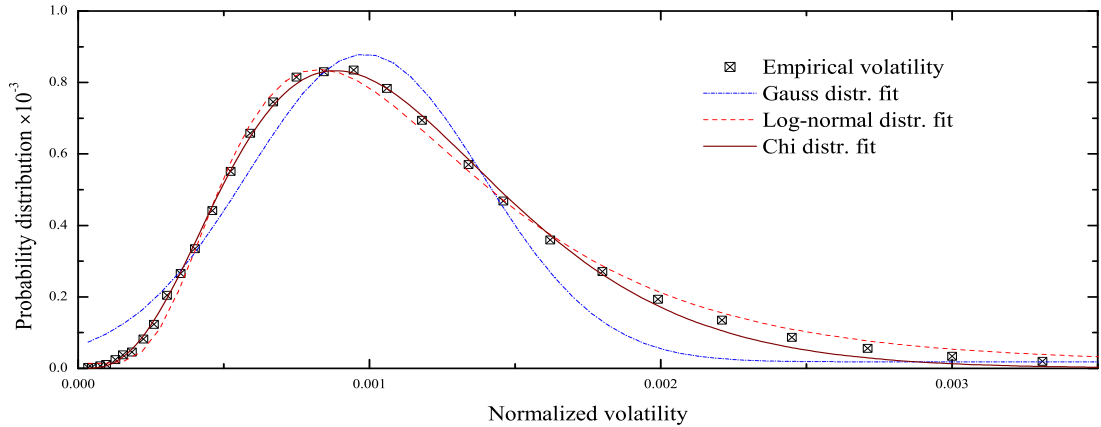}}}
\vspace{-0.8cm}\caption{\label{G.22b} Best Gaussian, Log-normal and
Chi-distribution fit for the S\&P 500 empirical volatility. The time
window $T = 300~$min.} \vspace{0.2cm} \hrule
\end{figure}

\begin{figure}[h]
\hspace{-0.4cm}\centerline{\resizebox{16.7cm}{7.4cm}{\includegraphics{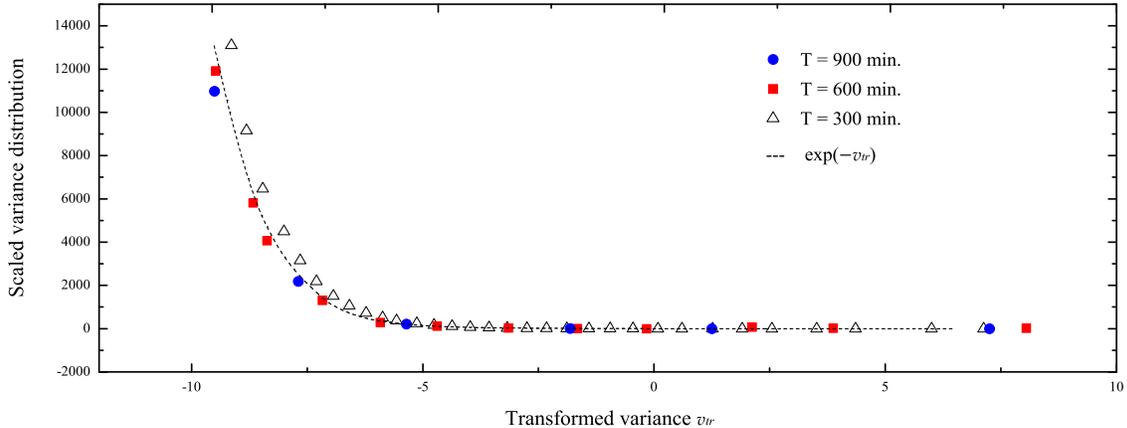}}}
\vspace{-0.8cm}\caption{\label{G.23} The scaled variance
distribution for different window sizes $T$. The scaled distribution
is plotted as a function of the transformed volatility $v_{tr} =
T\mu v + (1-T\nu) \ln (T \mu v)$.} \vspace{0.2cm} \hrule
\end{figure}

There is, however, a logical caveat in the definition (\ref{II.5}).
The observed volatility fluctuations have a spuriously superposed
pattern caused by intra-day fluctuations. This is because over the
day, the market activity is large at the beginning and at the end,
but exhibits a broad minimum around a noon~\cite{admati88}. Since
the volatility is supposed to measure the magnitude of the market
activity, this intra-day pattern should be removed in order to avoid
false correlations. This can be remedied by considering the
normalized returns~\cite{Liu97}
 \begin{eqnarray}\nonumber \\[-2mm]
R_{\rm{nor}}(t_n)  \equiv  \frac{\ln [S(t_n + \Delta t
)/S(t_n)]}{\langle |\ln [S(t_n + \Delta t )/S(t_n)]| \rangle} \cong
 \frac{S(t_n + \Delta t ) - S(t_n)}{S(t_n)}\bigg/\!\left\langle
\left| \frac{S(t_n + \Delta t ) - S(t_n)}{S(t_n)}\right|
\right\rangle \! . \label{II.7}
\\[-2mm] \nonumber\end{eqnarray}
Here $\langle \ldots \rangle$ is the average over the whole sequence
of trading days, i.e.,
\begin{eqnarray}\nonumber \\[-3mm]
\langle |\ln [S(t_n + \Delta t )/S(t_n)]| \rangle \ \equiv \
\frac{1}{N_{\rm{td}}}\sum_{j=1}^{N_{\rm{td}}} |\ln [S_j(t_n + \Delta
t )/S_j(t_n)]|\, ,
\\[-3mm] \nonumber\end{eqnarray}
with $S_j(t_n)$ denoting the spot price of the index at the time
$t_n$ on the $j$th trading day, and $N_{\rm{td}}$ denotes the total
number of trading days. So in (\ref{II.7}) each spot return
is divided by its natural ``spot return scale". The corresponding
normalized volatility is then defined as
\begin{eqnarray}\nonumber \\[-3mm]
\sigma_{T\rm{nor}}(t_{\eta}) \ \equiv \ \frac{1}{n} \sum_{m =
\eta}^{n-1 + \eta} |R_{\rm{nor}}(t_m)|\, . \label{II.6}
\\[-3mm] \nonumber\end{eqnarray}
This is the quantity whose distribution will be discussed here
(we omit the subscript ``nor" in the following). To
this end we shall examine the data set for the S\&P 500 stock market
index. Data in question were gathered over the period of 22 years
from Jan 1985 to Jan 2007 at roughly $5$ minute increments. The
corresponding empirical time sequence is seen in
Fig.~\ref{SP500index}.
With the help of the prescription (\ref{II.5}) we obtain the
volatility for the above S\&P 500 index data shown in
Fig.~\ref{SP500vol-notnorm}.
The corresponding normalized volatility $\sigma_{T\rm{nor}}(t)$ is
shown in Fig.~\ref{SP500vol}. The normalization was taken with
respect to $N_{\rm{td}} = 5550$ (i.e., approximate number of trading days
between 5 Jan. 1985 - 5 Jan. 2007).
Figure~\ref{G.22b} shows the volatility probability density function
$\rho_T(\sigma)$ with the time window $T = 300~$min. While it is
seen that $\rho_T(\sigma)$ can be well fitted with the Log-normal
function (as proposed in Ref.\cite{Liu97}), the Chi distribution
gives a better fit in the central part and is of  roughly the same
quality in the tail part. Because the value of the volatility
quantifies the asset risk, Fig.~\ref{G.22b} implies that the
Log-normal fit slightly overestimates, while the Chi distribution
slightly underestimates the large risks. In addition, the functions
$\mu$ and $\nu$ that parameterize the Chi (and Gamma) distribution
(cf. Eq.(\ref{Chi})) are in the observed time windows proportional
to $T$, i.e. $\mu(T) = \mu T$ and $\nu(T) = \nu T$. This tendency is
confirmed in Fig.~\ref{G.23} which shows the corresponding variance
distribution $\rho_T(v)$ for
three different window sizes $T$. For better comparison we use the
scaled distribution form, $\Gamma(T\nu)\rho_T(v)/(T\mu)$ as a
function of $T\mu v + (1-T\nu) \ln (T \mu v)$, with $\mu =
E(v)/Var(v)$ and $\nu = E^2(v)/Var(v)$ ($E$ and
$Var$ are empirical mean and variance, respectively). After the
above scaling, empirical variance distributions, with $T = 300,
600$ and $900~$min. nicely ``collapse" to the canonical exponential
$\exp(-x)$ (i.e., Gamma distribution with both mean and variance
equal to $1$), confirming thus the assumed Gamma distribution
behavior for rather long averaging times.
In the limit of very long $T$, one expects that $\rho_T(\sigma)$ becomes Gaussian, due to CLT.
For the times considered here, however, the Chi distribution fits the data
better than a Gaussian one. We see thus that the long-range correlations in the volatility
fluctuations considerably slow down the convergence towards the
Gaussian distribution. The conclusions that one my draw from this is
that, even for large times, the tail of $\rho_T(\sigma)$ has a
fairly large amount of distribution (or information) that cannot be
ignored. This will provide a key input in generalizing the
Black--Scholes pricing formula in Section~\ref{opf}.

\section{Theoretical digression -- Tsallis' density operator}
\label{SEc3}

We have mentioned before the result in statistical physics~\cite{kubo} that
a density matrix of a system with fluctuating temperature can be
written as a density matrix with fixed temperature averaged with
respect to some temperature distribution function. This technique was recenly
proposed as a natural frame for composing non-Gaussian price distributions~\cite{REMPI0}
and for deriving option pricing formulas~\cite{REPI}.

 Consider the Gaussian distribution of the Brownian motion of a particle of
unit mass as a function of the inverse temperature $\beta_G \equiv1/T$:
 \begin{eqnarray}\nonumber \\[-3mm]
 \rho_{\rm G}(x_b,x_a;\beta_G)
= \frac{1}{ \sqrt{2\pi\beta_G}}\ e^{-(x_b-x_a)^2/2 \beta_G }\, .
\\[-4mm] \nonumber\end{eqnarray}
We use natural units in which the Boltzmann constant $k_B$
has the value 1. The variance
$v\equiv  \sigma ^2$ of this distribution is obviously equal to $
\beta_G$. Following Ref.~\cite{REMPI0} we form a superposition of these distributions
as an integral over different inverse temperatures, i.e., different variances $v $:
\begin{eqnarray}\nonumber \\[-3mm]
 \rho_  \delta  (x_b,x_a;\beta) ~&=& \
\int_{0}^{\infty} \d v ~f_{\mu,
1/ \delta }(v)
 \frac{1}{ \sqrt{2\pi v}}\  e^{-(x_b-x_a)^2/2 v}, ~~~\;\;\;\;
 \beta \ \equiv\ \bar v \ = \ \nu /\mu\, ,
 \label{3.2}
\\[-3mm] \nonumber\end{eqnarray}
where $f_{\mu, \nu}(v) $ is the Gamma distribution defined in
(\ref{Gammad}) whose average lies at $\bar v = \nu /\mu$.
\comment{ If the Hamiltonian $H$ has the
standard form $p^2/2m + V(\bf{x})$, the path integral over momentum
can be easily done leaving a pure configuration-space path integral.
In particular, when the inverse temperature $\beta$ follows a Gamma
distribution, then the density matrix has the path-integral
representation\footnote{Here, and in subsequent integrations,
we ignore overall multiplicative factors.}
 \begin{eqnarray}\nonumber \\[-3mm]
 \rho(x_b,x_a;\beta) ~&=& \
\int_{0}^{\infty} \d \beta ~f_{\mu,
\nu}(\beta)\int_{x(0) = x_a}^{x(\beta) =x_b} {\mathcal{D}}x
\int\! {\mathcal{D}}p ~e^{-\int_{0}^{\beta} \d \tau (p\dot{x} -
H)}\, ,
 \label{3.2}
\end{eqnarray}
}%
The subscript $\delta $ of the distribution characterizes the ratio
\begin{eqnarray}\nonumber \\[-3mm]
 \delta =\frac{1} \nu =\frac{\overline{(v-\bar v)^2}}{\bar v^2},
\\[-3mm] \nonumber\end{eqnarray}
a quantity which we shall call the {\em spread\/} of the Gamma
distribution. In terms of the integration variable $s=\mu v $,
we can rewrite (\ref{3.2}) in the form
\begin{eqnarray}\nonumber \\[-3mm]
\rho_{\delta} (x_b,x_a;\beta) \ = \ \frac{1}{\Gamma(1/\delta)}
\int_{0}^{\infty} \frac{\d s}{s} ~s^{1/\delta} e^{-s}
\sqrt{\frac{\mu}{2\pi s}} \ e^{-(x_b-x_a)^2\mu/2s }\, .
 \label{eq2.1}
\\[-3mm] \nonumber\end{eqnarray}
In quantum mechanics, one writes such a distribution
in a notation due to Dirac as a matrix element
\begin{eqnarray}
\rho(x_b,x_a;\beta)\ \equiv \ \langle x_b|\hat{\rho}(\beta)|x_a
\rangle\, ,
\\[-3mm] \nonumber\end{eqnarray}
of a density  operator
%
 \begin{eqnarray}\nonumber \\[-3mm]
\hat{\rho}(\beta) ~\equiv ~\frac{1}{\Gamma(1/\delta)}
\int_{0}^{\infty} \frac{\d s}{s} ~s^{1/\delta} e^{-s} e^{-\delta
\beta\, s \hat{H} }\, ,\label{3.3}
~~~~~\beta\equiv
 \nu /\mu=1/\mu \delta \, ,
\\[-3mm] \nonumber\end{eqnarray}
where
 \begin{eqnarray}
\hat H \ \equiv \ \frac{\hat p^2}{2} \ = \ -\frac{
\partial _x^2}{2}\, ,
\label{@HamIl}\\[-3mm] \nonumber\end{eqnarray}
is the Hamilton operator of a free particle of unit mass. The
integral over $s$  can be done (as in the so-called  Schwinger
trick~\cite{HK}) and yields:
\begin{eqnarray}\nonumber \\[-3mm]
\hat{\rho}( \beta) ~= ~\left[ 1  \ + \
 \beta  \delta  \hat{H}
\right]^{- 1/ \delta }\, . \label{3.1}
\\[-3mm] \nonumber\end{eqnarray}
This is the (un-normalized) Tsallis density operator (cf. Appendix~B)
with a so-called escort parameter $q$ related to the spread
parameter $\delta$ by $q \equiv 1 + \delta$, and an inverse
Tsallis temperature $\beta$
equal to  $\nu/\mu$.
%
%
%
In the limit $\delta \rightarrow 0$ where $q\rightarrow 1_+$, the
Tsallis operator (\ref{3.1}) converges to an exponential
\begin{eqnarray}
\hat{\rho}( \beta) \ \rightarrow \ e^{- \beta_G
\hat{H}},\;\;\;\;\;\;\;\; \beta_G \ = \ \beta|_{q=1}\, ,
\label{3.1p}
\\[-3mm] \nonumber\end{eqnarray}
and the distribution function (\ref{eq2.1}) becomes Gaussian.
The corresponding limit
of the Gamma distribution (\ref{Gammad}) is a
$ \delta $-function:
 \begin{eqnarray}\nonumber \\[-3mm]
f_{\mu, \nu}(x) \approx  \sqrt{\frac{\mu}{2\pi x}} \!
 \left(\frac{\mu x}{\nu}
 \right)^{\nu -1/2} e^{-\nu(\mu x/\nu -1 )}  \approx
 \frac{1}{\sqrt{2}~x} ~\delta\!\left( \sqrt{ \frac{\mu x}{\nu} -1 - \log
 \frac{\mu
 x}{\nu}}\right)  =  \delta\!\left(x - \frac{\nu}{\mu}\right)\!
 .
\label{@LiMit} \\[-1mm]\nonumber\end{eqnarray}
Hence the density matrix (\ref{3.2}) approaches for $ \delta
\rightarrow 0$ the canonical density matrix of the Gibbs-Boltzmann
statistics of the inverse temperature $\nu/\mu = \beta_G$. Note,
that in order to ensure that $\beta_G$ is finite in the
small-$\delta$ limit, $\mu$ must behave as $1/\delta \beta_{G}$ at
$\delta \rightarrow 0$.  The reader may easily check that a
superposition of the type (\ref{3.2}) exists also for distributions
of Bashkirov's 1-st version of thermostatics (cf. Appendix~B). In
that case $0 < q< 1$, $\delta = 1-q$ and $\beta \mapsto
\tilde{\beta}$. In the limit $q\rightarrow 1_-$ the density matrix
becomes again a Gibbs-Boltzmann canonical density matrix.

\section{Generalized option pricing formula} \label{opf}

Consider a continuously tradable stock and assume that the stock
fluctuations over short time intervals $ \Delta t_0$  such as $
\Delta t_0=1$ day are described by the stationary Tsallis density
operator (\ref{3.1}). The time $ \Delta t_0$ will from now on play
the role of a time unit. We now analyze the modifications of the
Black-Scholes formula for pricing European call options \cite{CallE}
brought about by the fluctuations of the volatilities. After a time
$t>0$ (always in units of $ \Delta t_0$) the stock fluctuations
follow a Tsallis density operator
 \begin{eqnarray}\nonumber \\[-3mm]
[\hat{\rho}( \beta)]^t ~&=& ~Z^{-t}\left[1 ~+
 ~ \beta \delta \hat{H}\right]^{-t/\delta}\nonumber ~= \
 \frac{Z^{-t}}{\Gamma\left(t/\delta\right)}\int_{0}^{\infty}\frac{\d
 s}{s}~s^{t/\delta}e^{-s[1 +  \beta \delta \hat{H}]}
 \nonumber \\[3mm]
&=&
 ~\left(\frac{t}{ \beta \delta}\right)^{\! t/\delta}
 \frac{Z^{-t}}{\Gamma\left(t/\delta\right)}\int_{0}^{\infty}\frac{\d
 v}{v}~v^{t/\delta}e^{-tv\mu}e^{-tv\hat{H}}\, , \label{cd.1}
\\[-1mm] \nonumber\end{eqnarray}
%
where $Z$ is a normalization factor. In the sequel we shall allow
for a drift of the returns by extending the Hamiltonian operator
(\ref{@HamIl}) to
\begin{eqnarray}
H = p^2/2 + p r_{x_W}/v\, , \label{HamI2}\end{eqnarray}
where $ r_{W}$ is the growth rate of the riskfree investment, and
$r_{x_W}= r_{W} + v/2$ is the associated growth rate of its
logarithm. The parameter $ \beta $ equals to $1/ \delta \mu$, as
before. The matrix elements of (\ref{cd.1}) yield the
time-compounded probability density
\begin{eqnarray}\nonumber \\[-3mm]
P_\delta(x_b,t_b;x_a,t_a)~= ~\langle x_b|[\hat{\rho}( \beta)]^t
|x_a\rangle\, ,  \;\;\;     ~~~~t \equiv t_b - t_a >0\, .
\label{IV9b}
\\[-3mm] \nonumber\end{eqnarray}
\comment{can be conveniently rewritten in the path-integral form
 \begin{eqnarray}\nonumber \\[-3mm]
P_{\delta}(x_b,t_b;x_a,t_a) ~= ~Z^{-t}\int_{0}^{\infty}\!\! \d v \
f_{t\mu,~\!t/\delta}(v)\int^{x(t_b) = x_b}_{x(t_a)=
x_a}{\mathcal{D}}x \int{\mathcal{D}}p ~e^{- \int^{t_b}_{t_a}\d
\tau(p\dot{x} - v H)}. \label{IV1}
\\[-1mm] \nonumber\end{eqnarray}
} Being the matrix element of the product of operators $\hat{\rho}(
\beta)$. $P_\delta(x_b,t_b;x_a,t_a)$ fulfills trivially the
Chapman-Kolmogorov relation for a Markovian process
 \begin{eqnarray}\nonumber \\[-3mm]
P_{\delta}(x_b,t_b;x_a,t_a) \ = \ \int_{-\infty}^{\infty}\d x \
P_{\delta}(x_b,t_b;x,t_c) P_{\delta}(x,t_c;x_a,t_a)\, , \;\;\;\;\;
t_b > t_c > t_a\, . \label{IV1aa}
\\[-3mm] \nonumber\end{eqnarray}

\begin{figure}[h]
\centerline{\resizebox{18cm}{5,5cm}{\includegraphics{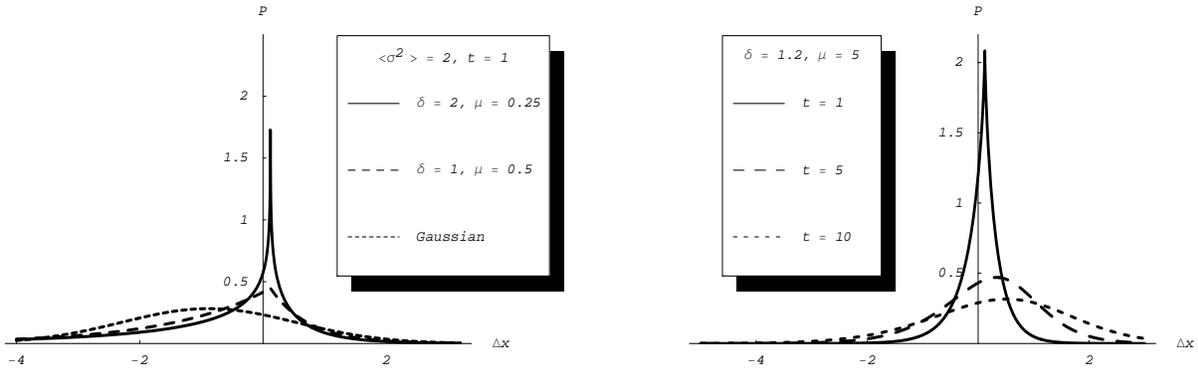}}}
\vspace{-0.6cm}\caption{\label{G.2} Normalized measure density
(\ref{IV2}); for various values of $\mu$ and $\delta$ at fixed
expiration time and $\langle \sigma^2 \rangle$ (left), for different
expiration times at fixed $\delta$ and $\mu$ (right). We set the
riskless interest rate $r_W$ to be $12\%$.} \vspace{0.2cm} \hrule
\end{figure}

For Gaussian stock fluctuations with the Hamiltonian (\ref{HamI2}),
the riskfree martingale measure density has the form~\cite{HK}
 \begin{eqnarray}\nonumber \\[-3mm]
&&P^{(M,r_W)}_ v (x_b,t_b;x_a,t_a) = \Theta(t_b -
t_a)\frac{e^{-r_W(t_b - t_a)}}{\sqrt{2\pi v(t_b - t_a)}}\
\exp\left\{- \frac{[x_b - x_a - r_{x_W}(t_b - t_a)]^2}{2 v(t_b -
t_a)}
 \right\}\nonumber \\[3mm]
&&\mbox{\hspace{2cm}}= \Theta(t_b - t_a) e^{-r_W(t_b - t_a)}
\int_{x(t_a) = x_a}^{x(t_b) = x_b}{\mathcal{D}}x ~\exp\left\{-
\frac{1}{2 v}\int_{t_a}^{t_b}[\dot{x} - r_{x_W}]^2 \d t \right\}.
 \label{IV1a}
\\[-1mm] \nonumber\end{eqnarray}
The corresponding time-compounded version of the superposition (\ref{3.2})
can be directly written as
 \begin{eqnarray}\nonumber \\[-3mm]
&&P_ \delta (x_b,t_b;x_a,t_a) ~= ~ \int_{0}^{\infty}\!\! \d v
~f_{t\mu,\ \!t/\delta}(v)\!~P_v^{(M,r_W)}(x_b,t_b;x_a,t_a)\, .
\end{eqnarray}
Correctness of the latter can be verified by combining (\ref{cd.1}),
(\ref{IV9b}), and (\ref{IV1a}). The result is \cite{REMPI0}
 \begin{eqnarray}\nonumber \\[-3mm]
&&P_\delta(x_b,t_b;x_a,t_a) ~= ~ \int_{0}^{\infty}\!\! \d v   ~f_{t\mu,\
\!t/\delta}(v)\!~P_v^{(M,r_W)}(x_b,t_b;x_a,t_a)\nonumber\\[3mm]
&&\mbox{\hspace{0.5cm}} = ~ \frac{e^{- (r_W t + \Delta x)/2}(2\mu
)^{t/\delta}
     }{\sqrt{\pi }\,\Gamma\left(t/\delta\right)} \! \left(\frac{\sqrt{1 +
8\,\mu}}{|\Delta x - r_W t|}\right)^{\!\!1/2 - t/\delta}\!\! K_{1/2 -
t/\delta}\!\!~\left(|\Delta x - r_W
     t|~\frac{\sqrt{1 + 8\,\mu }}{2}\right)\!
, \label{IV2}
\\[-1mm] \nonumber\end{eqnarray}
with $\Delta x \equiv x_b - x_a$ and $t_b >t_a$. The measure density
(\ref{IV2}) has for $|\Delta x| \gg 1$ the asymptotic behavior of the Erlang
 distribution~\cite{feller}:
 \begin{eqnarray}\nonumber \\[-3mm]
P_\delta(x_b,t_b;x_a,t_a) ~\approx ~\exp\left(-\frac{|\Delta x|}{2}\sqrt{1 +
8\,\mu } - \frac{\Delta x}{2}\right) ~|\Delta x|^{-1 + (t_b-t_a)/\delta }\, ,
 \label{IV2b}
\\[-1mm] \nonumber\end{eqnarray}
which has a semi-fat tail (see Fig.~\ref{G.2}). It should be noted
that the exponential suppression ensures that all momenta are
finite. This fact is an important ingredient in showing that
(\ref{IV2}) represents indeed the riskfree martingale measure
density. The actual proof of the latter can be found in
Ref.~\cite{HK}.

From the martingale measure (\ref{IV2}), we now calculate the option
price at an arbitrary {\em earlier} time $t_a$ via the evolution
equation
\begin{eqnarray}\nonumber \\[-3mm]
 O_ \delta (x_a,t_a) ~= ~\int_{-\infty}^{\infty}\d x_b ~O(x_b,t_b)P_\delta(x_b,t_b;x_a,t_a)
 \, .
\label{IV3}
\\[-2mm] \nonumber\end{eqnarray}
The value of the option at its expiration date $t_b$ is given by
the difference between the underlying stock price  $S_b \equiv
S(t_b)$ on expiration date and the strike price $E$, i.e.,
 \begin{eqnarray}\nonumber \\[-3mm]
O(x_b,t_b)~= ~\Theta(S_b - E)(S_b - E) ~= ~\Theta(x_b -
x_E)\left(e^{x_b} - e^{x_E}\right)\, , \label{IV4}
\\[-4mm] \nonumber\end{eqnarray}
where $x_E \equiv \log E$. Heaviside function $\Theta(x)$ ensures
that the owner of the option will exercise his right to buy the
stock only if he profits, i.e., only when $S_b - E$ is positive.

For  Gaussian fluctuations of variance $v$, formula (\ref{IV3}) is
evaluated with the measure (\ref{IV1a}) and yields directly the
Black-Scholes option pricing formula~\cite{B-P}:
 \begin{eqnarray}\nonumber \\[-3mm]
O_v^{(BS)}(x_a,t_a) ~= ~S(t_a)\Phi(y^+_v) ~- ~e^{-r_W(t_b - t_a)}E
\Phi(y^-_v)\, ,\label{IV5}
\\[-3mm] \nonumber\end{eqnarray}
where $\Phi(y)$ represents the cumulative normal distribution
and
\begin{eqnarray}\nonumber \\[-3mm]
y^{\pm}_v ~= ~\frac{\log[S(t_a)/E] ~+ ~\left(r_W \pm
\mbox{$\frac{1}{2}$}v \right)(t_b - t_a)}{\sqrt{v(t_b
- t_a)}}\, \label{IV6} .
\\[-1mm] \nonumber\end{eqnarray}
The generalization to the present semi-heavy tail distribution
is obtained from the superposition~\cite{REMPI0}
 \begin{eqnarray}\nonumber \\[-3mm]
O_ \delta (x_a,t_a) ~= ~ \int_{0}^{\infty}\!\! \d v   ~f_{t\mu,\
\!t/\delta}(v) ~O^{(BS)}_v(x_a,t_a)\, , \;\;\;\;\;\;\;\;\;\;\; t
> 0\, . \label{IV7}
\\[-2mm] \nonumber\end{eqnarray}
At this stage one should realize that integration in (\ref{IV7})
acts only on the $\Phi(y^{\pm}_v)$ parts of $O^{(BS)}_v(x_a,t_a)$.
This allows to write the option price in the Black-Scholes-like
form, namely
 \begin{eqnarray}\nonumber \\[-3mm]
O_{\delta}(x_a,t_a) ~ = ~S(t_a)\Phi^{(+)}_v ~- ~e^{-r_W(t_b - t_a)}E \Phi^{(-)}_v\, ,
\label{IV8}
\\[-4mm] \nonumber\end{eqnarray}
where $\Phi^{(+)}_v$ and $\Phi^{(-)}_v$ are Gamma-smeared versions
of the functions $\Phi(y^+_v)$ and $\Phi(y_v^-)$, respectively. For
$\delta \rightarrow 0$, the new functions reduce, of course, to the
un-smeared ones due to the limit (\ref{@LiMit}).

Let us now do the smearing operation for $\Phi^{(+)}_v$:
 \begin{eqnarray}\nonumber \\[-4mm]
\Phi^{(+)}_v ~&=& ~\int_{0}^{\infty}\d v~f_{t\mu, t/\delta}(v) \Phi(y_v^+) \
= ~\int_{0}^{\infty}\d v~f_{t\mu, t/\delta}(v)
\int_{-\infty}^{y_v^+}\frac{\d \xi}{\sqrt{2\pi}}\
e^{-\xi^2/2}\nonumber \\[2mm]
&=& ~\int_{0}^{\infty}\d v~f_{t\mu, t/\delta}(v)
\int_{-\infty}^{\infty}\frac{\d \xi}{\sqrt{2\pi}}~\Theta (y_v^+ -
\xi) ~e^{-\xi^2/2} .
\end{eqnarray}
By expressing the Heaviside function as
\begin{eqnarray}
\Theta(x)=\int _{-\infty}^\infty \frac{\d p}{2\pi i}\ \frac{1}{p-i0_+} \ e^{ipx},
\label{@}\end{eqnarray}
and using Sokhotsky's formula
 \begin{eqnarray}\nonumber \\[-4mm]
\frac{1}{p-i0_+} ~= ~{\cal{P}}\!\left(\frac{1}{p}\right) ~+ \
i\pi \delta(p)\, ,
\\[-2mm] \nonumber\end{eqnarray}
where ${\cal P}$ denotes the principal value of the integral,
we perform the integral over $\xi$ and obtain~\cite{SEE}
 \begin{eqnarray}
\Phi^{(+)}_v ~&=&
 ~\frac{1}{2} ~+ ~{{\cal P}}\int_{-\infty}^{\infty}\frac{\d
p}{2\pi i} ~F^{(+)}_{t\mu, t/\delta}(p,A) ~ \frac{e^{-p^2/2}}{p} \,
.\label{IV9}
\\[-2mm] \nonumber\end{eqnarray}
Here $F^{(+)}_{t\mu, t/\delta}(p,A)$ is the integral
\begin{eqnarray}\nonumber \\[-4mm]
F^{(+)}_{t\mu, t/\delta}(p,A)\ \equiv \ \int_{0}^{\infty}\d v
~f_{t\mu, t/\delta}(v) \ e^{ipy_v^+}\, .\label{IV99}
\\[-2mm] \nonumber\end{eqnarray}
This is evaluated as follows. We perform a change of variables from
$v$ to $\omega\equiv  \sqrt{vt}$, and define $A \equiv \log [S(t_a)/E] + r_W t$,
then we  have
 \begin{eqnarray}\nonumber \\[-4mm]
F^{(+)}_{t\mu, t/\delta}(p,A)&=& 2
\frac{\mu^{t/\delta}}{\Gamma(t/\delta)} \int_{0}^{\infty}\d \omega \
\omega^{2t/\delta -1} ~e^{-\mu \omega^2} \
\exp\left[i\frac{p}{2}\left(\omega + \frac{2A}{\omega}
\right)\right] \nonumber \\[2mm]  &\equiv& ~2
\frac{\mu^{t/\delta}}{\Gamma(t/\delta)} ~g^{(+)}_\mu(p,A)~.\label{IV11}
\\[-2mm] \nonumber\end{eqnarray}
In order to calculate $g^{(+)}_\mu(p,A)$ we observe that the
following integral can immediately be done:
\begin{eqnarray}\nonumber \\[-4mm]
\int_{0}^{\infty}\d \mu ~\mu^{s-1}g^{(+)}_\mu(p,A) ~= ~\Gamma(s)
\int_{0}^{\infty}\d \omega~\omega^{2t/\delta -2s -1}\
\exp\left[i\frac{p}{2}\left(\omega + \frac{2A}{\omega}
\right)\right] \, .
\\[-2mm] \nonumber
\end{eqnarray}
Let us first assume that $p>0$ and Arg$(A) =0$ and perform the
substitution $\omega = \sqrt{2A}~\!\exp u$. This yields directly
 \begin{eqnarray}\nonumber \\[-4mm]
\int_{0}^{\infty}\d \mu ~\mu^{s-1}g^{(+)}_\mu(p,A) ~&=& \
\Gamma(s)(2A)^{t/\delta - s} \int_{-\infty}^{\infty}\d u \
e^{ip\sqrt{2A}\cosh u - (2s - 2t/\delta)u}\nonumber \\[2mm]
&=& ~i\pi \Gamma(s)(e^{-i\pi} 2A)^{t/\delta - s } H^{(1)}_{2s -
2t/\delta}(p\sqrt{2A})\, , \;\;\;\;\;\;\;\;\;\;\;\;\;\; (p>0)\, ,
\\[-2mm] \nonumber\end{eqnarray}
where $H^{(1)}_{ \alpha }(z)$ is the Hankel function of the first
kind~\cite{Watson}. This result holds for $\Re(s) > t/\delta - 1/2$.
Similarly we obtain for $p<0$:
 \begin{eqnarray}\nonumber \\[-4mm]
\int_{0}^{\infty}\d \mu ~\mu^{s-1}g^{(+)}_\mu(p,A) ~&=& ~-i\pi
\Gamma(s)(e^{i\pi} 2A)^{t/\delta - s } H^{(2)}_{2s -
2t/\delta}(-p\sqrt{2A}) \, , \;\;\;\;\;\;\;\;\;\;\;\;\;\; (p<0)\, ,
\\[-2mm] \nonumber\end{eqnarray}
where $H^{(2)}_{ \alpha }(z)$ is the Hankel function of the second
kind~\cite{Watson}. The result is valid for $\Re(s) < t/\delta+ 1/2$.

Knowing these integrals, we find the function $g^{(+)}_\mu(p,A)$
itself with the help of the Mellin inverse transform:
 \begin{eqnarray}\nonumber \\[-2mm]
g^{(+)}_\mu(p,A)= \int_{c-i\infty}^{c+i\infty}\frac{\d s}{2\pi i} \
\mu^{-s}\!\left\{
          \begin{array}{l}
            ~~\,i\pi \Gamma(s)(e^{-i\pi} 2A)^{t/\delta - s } H^{(1)}_{2s -
2t/\delta}(p\sqrt{2A}),~~~~p>0,\\[2mm]
            -i\pi
\Gamma(s)(e^{i\pi} 2A)^{t/\delta - s } H^{(2)}_{2s -
2t/\delta}(-p\sqrt{2A}),  ~~~~p<0.
          \end{array}
        \right\}
\\[-2mm] \nonumber\end{eqnarray}
where $c \in (t/\delta - 1/2, t/\delta+ 1/2)$.
Inserting this back into (\ref{IV11}) and (\ref{IV9}) one can
write for $\Phi^{(+)}_v$:
 \begin{eqnarray}\nonumber \\[-2mm]
\!\!\!\!\!\!\!\!\!\!\!
\!\!\!\!\!\!\!\!\!\!\!
\!\!\!\!\!\!\!\!\!\!\!
\!\!\!\!\!\!\!\!\!\!\!
\!\!\!\!\!\!\!\!\!\!\!\Phi^{(+)}_v  &=&  \frac{1}{2} + \frac{\mu^{t/\delta}}{\Gamma(t/\delta)}\
\!{{\cal P}}\!\!\int_{-\infty}^{\infty} \frac{\d p}{2\pi i}
\frac{e^{-p^2/2}}{p} \int_{c-i\infty}^{c + i\infty}\!\!\d s\
\Gamma(s)\mu^{-s} \! \left\{
                      \begin{array}{l}
                        (e^{-i\pi} 2A)^{t/\delta - s } H^{(1)}_{2s -
2t/\delta}(p\sqrt{2A})\!\!\!\!\!\!\!\!\!\!\!\!\!\! \\[2mm]
                        -(e^{i\pi} 2A)^{t/\delta - s } H^{(2)}_{2s -
2t/\delta}(-p\sqrt{2A})
                      \end{array}
                    \right\} \nonumber\!\!\!\!\!\!\! \\[2mm]
&=&  \frac{1}{2} +
\frac{\mu^{t/\delta}}{\Gamma(t/\delta)}\int_{c-i\infty}^{c +
i\infty}\!\!\d s~\Gamma(s) \mu^{-s} (2A)^{t/\delta -s}
\int_{0_+}^{\infty} \frac{\d p}{2\pi i} \frac{e^{-p^2/2}}{p}
\!\!\!\!\!\!\!\!\!\!\!\!\!\!\!\!\!\!\!\!\!\!\!\!\!\!\!\nonumber \\[1mm]
&&\mbox{\hspace{2cm}}\times  \left[e^{i\pi(s -t/\delta )} H_{2s -
2t/\delta}^{(1)}(p\sqrt{2A}) + e^{i\pi(t/\delta - s)} H_{2s -
2t/\delta}^{(2)}(p\sqrt{2A})\right].
\!\!\!\!\!\!\!\!\!\!\!\! \!\!\!\!\!
\!\!\!\!\!\!\!\!\!\!\!\! \!\!\!\!\!
\!\!\!\!\!\!\!\!\!\!\!\! \!\!\!\!\!
\\[-2mm] \nonumber\end{eqnarray}
Decomposing the Hankel functions into Bessel functions of first
kind,
 \begin{eqnarray}\nonumber \\[-3mm]
H_ \nu^{(1,2)} (z)\ \equiv \ \pm \frac{J_{-\nu}(z) \ - \ e^{\mp
i\pi\nu}J_{\nu}(z)}{i \sin(\nu \pi)}\, ,
\end{eqnarray}
this becomes
 \begin{eqnarray}\nonumber \\[-2mm]
\!\!\!\!\!\!\!\!\!\!\!
\!\!\!\!\!\!\!\!\!\!\!
\Phi^{(+)}_v
&=& \frac{1}{2} + \frac{(2A\mu)^{t/\delta}}{\Gamma(t/\delta)}~\!
\int_{c-i\infty}^{c + i\infty}\!\!~\frac{\d s}{2\pi i} ~\frac{\
\Gamma(s)
(2A\mu)^{-s}}{\cos[\pi(t/\delta -s)]}~\! \int_{0_+}^{\infty} \d p ~\frac{e^{-p^2/2}}{p}~\nonumber \\[1mm]
&&\mbox{\hspace{4cm}}\times  \left[J_{2s - 2t/\delta}(p\sqrt{2A}) +
J_{ 2t/\delta - 2s}(p\sqrt{2A})\right].\!\!\!\!\!\!\!\!\!\!
\label{@CONT}
 \\[-2mm] \nonumber\end{eqnarray}
The $p$-integral can now be easily performed yielding
 \begin{eqnarray}\nonumber \\[-2mm]
&&\frac{1}{\cos(\pi\zeta)}\int_{0_+}^{\infty} \d p \
\frac{e^{-p^2/2}}{p}\left\{
                      \begin{array}{l}
                        J_{2\zeta}(p\sqrt{2A}) \\
                        J_{
-2\zeta}(p\sqrt{2A})
                      \end{array}
                    \right\}\nonumber \\[2mm]
&&\mbox{\hspace{0.6cm}}= ~\frac{1}{2\sqrt{\pi}\zeta} \left\{
  \begin{array}{ll}
   \!\left(\frac{A}{4}\right)^{\zeta}\Gamma\!\left(\mbox{
$\!\!\frac{1}{2}$} -\zeta\right)\!\!~_{1\!}F_1(\zeta, 1 + 2\zeta,
-A), & \hbox{~~~$\Re~\zeta > 0$} \\[2mm]
   \!- \left(\frac{A}{4} \right)^{-\zeta}\Gamma\!\left(
 \mbox{
$\!\!\frac{1}{2}$}+\zeta
\right)\!\!~_{1\!}F_1(-\zeta, 1- 2\zeta, -A), &
\hbox{~~~$\Re~\zeta < 0$}
  \end{array}
\right\} ,
\\[-2mm] \nonumber\end{eqnarray}
where $\zeta \equiv s-t/\delta$. The function
$\!\!\!~_{1\!}F_1(a,b,z)$ is the Kummer confluent hypergeometric
function~\cite{Buchholtz}. Note also that the fundamental strip
$c\in (t/\delta - 1/2, t/\delta+ 1/2)$ is for $\zeta$ in the upper
expression reduced to $c \in (0, t/\delta + 1/2)$, while for the
lower expression $c \in (t/\delta -1/2,0)$.

We now perform  $s$-integral in (\ref{@CONT}), and obtain
 \begin{eqnarray}\nonumber \\[-4mm]
\!\!\!\!\!\!\Phi^{(+)}_v ~=~\frac{1}{2} ~+ ~h^{(+)}_1(\mu,t/\delta,
A) ~- ~h^{(+)}_2(\mu, t/\delta, A)\, ,\label{phi-plus}
\end{eqnarray}
with
 \begin{eqnarray}\nonumber \\[-4mm]
&&\mbox{\hspace{-11mm}} h^{(+)}_1(\mu,t/\delta, A) ~= ~ \frac{1}{2\Gamma(t/\delta)\sqrt{\pi}
}~\sum_{\mathrm{Res}} \frac{\Gamma(\zeta +
t/\delta)}{\zeta}\left(8\mu\right)^{-\zeta}
 \Gamma\!\left(\mbox{
$\!\!\frac{1}{2}$} -\zeta\right) \!~_{1\!}F_1(\zeta, 1 + 2\zeta,
-A)\, ,\nonumber \\[2mm]
&&\mbox{\hspace{-11mm}} h^{(+)}_2(\mu,t/\delta, A) ~= ~ \frac{1}{2\Gamma(t/\delta)\sqrt{\pi}
}~\sum_{\mathrm{Res}} \frac{\Gamma(\zeta + t/\delta)}{{\zeta}}
\left(\frac{A^2\mu}{2}\right)^{-\zeta}\!\! \Gamma\!\left(\zeta + \mbox{
$\!\!\frac{1}{2}$}\right) \!~_{1\!}F_1(-\zeta, 1- 2\zeta, -A).
\\[-1mm] \nonumber\end{eqnarray}

To compute the residues of the poles of $h^{(+)}_1$ and $h^{(+)}_2$ we
need to decide in what way the poles are enclosed in the complex
plane. Taking into account Stirling's large-argument limit
of the Gamma functions:
\begin{eqnarray}\nonumber \\[-3mm]
\frac{\Gamma(\zeta + t/\delta)\Gamma(\mbox{ $\!\!\frac{1}{2}$}
-\zeta)}{\zeta} ~\approx ~\zeta^{t/\delta -3/2},
\;\;\;\;\;\mbox{}\;\;\; \frac{\Gamma(\zeta +
t/\delta)\Gamma(\mbox{ $\!\!\frac{1}{2}$} + \zeta)}{\zeta} ~\approx
~\left(\frac{\zeta}{e} \right)^{2\zeta}\zeta^{t/\delta -3/2}\, ,
\\[-3mm] \nonumber\end{eqnarray}
for $|\zeta|\rightarrow\infty$ ($|\mbox{Arg}(\zeta)| < \pi$),
and the fact that the asymptotic behavior of $~_{1\!}F_1$ for large
$|\zeta|$ is given by Kummer's second formula [see, e.g.,
Ref.~\cite{Luke} Eq. (4.8.16)]
 \begin{eqnarray}\nonumber \\[-3mm]
&&~_{1\!}F_1(\pm\zeta, 1\pm 2\zeta, -A) ~\approx ~e^{-A/2},
\;\;\;\;\; |\mbox{Arg}(\zeta)| ~<  ~\pi\, ,\nonumber \\[2mm]
&&~_{1\!}F_1(\zeta, 1 + 2\zeta, -A)\;\;\;\;\;\
\mbox{for}\;\;\;\;\;\; \zeta ~= ~-1,-2,-3, -4, \ \ldots \;\;\;\;
\mbox{undefined}\, ,
\\[-3mm] \nonumber\end{eqnarray}
we obtain that the contour closure of $h^{(+)}_2$ depends entirely on the
behavior of the Gamma functions. In fact, due to previous asymptotics
we must close the contour in $h^{(+)}_2$ to the left as the value of the
contour integral around the large arc is zero in the limit of
infinite radius. Due to $(8\mu)^{-\zeta}$ term $h^{(+)}_1$ closes the contour
to the right. There are only simple poles contributing
to $h^{(+)}_1$, which lie at $\zeta = (2n +1)/2, \; n  \in {\mathbb{N}}
\equiv (0,1,2,\dots)$. If we assume for a moment that
$t/\delta \not = 1/2 + l, \; l \in {\mathbb{N}}$, then the only
singularities of $h^{(+)}_2$ are due to
simple poles at $\zeta = -1/2 - m, \; m \in {\mathbb{N}}$ and $\zeta =
- t/\delta -k, \;  k\in {\mathbb{N}}$. Consequently we can write
 \begin{eqnarray}\nonumber \\[-2mm]
\Phi^{(+)}_v &=& \frac{1}{2}\left[1 ~+ ~\sqrt{\frac{
\bar v  \delta }{2\pi}}
\sum_{n=0}^{\infty}\frac{(t/\delta)_{n+1/2}(1/2)_n}{(1)_{2n+1}}\left(
-\frac{\bar v\delta}{2}\right)^{\!\!
n}\!~_{1\!}F_1\left(n+\frac{1}{2}, 2n + 2, -A\right) \right. \nonumber \\[2mm]
&&\mbox{\hspace{-15mm}}  + \left. ~A \sqrt{\frac{2}{\pi \bar
v\delta}}
\sum_{n=0}^{\infty}\frac{(t/\delta)_{-n-1/2}(1/2)_n}{(1)_{2n+1}}\left(
-\frac{2A^2}{\bar v\delta}\right)^{\!\!
n}\!~_{1\!}F_1\left(n+\frac{1}{2}, 2n + 2, -A\right)\right. \nonumber \\[2mm]
&&\mbox{\hspace{-15mm}}  + \left. \frac{1}{\cos(\pi
t/\delta)}\left(\frac{2A^2}{\delta \bar v }\right)^{\!\! t/\delta}
\sum_{n=0}^{\infty}\frac{(t/\delta)_n}{(1)_{2n +  2t/\delta} (1)_n}
\left(\frac{2A^2}{\delta \bar v} \right)^{\!\!n}\!~_{1\!}F_1\left(n+
\frac{t}{\delta}, 2n + 1 + 2\frac{t}{\delta}, -A\right) \right]\!,
\label{IV9c}
\\[-1mm] \nonumber\end{eqnarray}
where we have set
$\bar v\equiv \langle v \rangle=\langle  \sigma ^2\rangle$ and used
the Pochhammer symbols $(z)_{k}~\equiv ~{\Gamma(k+z)}/{\Gamma(z)}\, $.
If $ \delta $ is very small, or  $t$ very large,
we can use the asymptotic behavior
$(t/\delta)_{z} \rightarrow (t/\delta)^{z}$ and the fact that the
sum in the third line of (\ref{IV9c}) tends to zero due to
strong suppression by $1/(1)_{2n + t/\delta}$. In this case
(\ref{IV9c}) reduces to
\begin{eqnarray}\nonumber \\[-3mm]
\Phi^{(+)}_v ~= ~\frac{1}{2}\left[1 ~+ ~\sqrt{\frac{2}{\pi}}\
y_{\bar v}^+
 \!~_{1\!}F_1\left(\frac{1}{2}, \frac{3}{2}, -
\frac{(y_{\bar v}^+)^2}{2}\right) \right] ~= ~\Phi( y_{\bar v}^+) \,,~~~~\bar v= \beta\, .
\\[-3mm]\nonumber\end{eqnarray}
The calculation of $\Phi^{(-)}_v$ is similar and is relegated  to Appendix~A.
Here we state only the result:
 \begin{eqnarray}\nonumber \\[-3mm]
\Phi^{(-)}_v ~&=& ~\frac{1}{2}\left[1 ~- ~\sqrt{\frac{\delta \bar v }{2\pi}}
\sum_{n=0}^{\infty}\frac{(t/\delta)_{n+1/2}(1/2)_n}{(1)_{2n+1}}\left(
-\frac{\bar v\delta}{2}\right)^{\!\!
n}\!~_{1\!}F_1\left(n+\frac{1}{2}, 2n + 2, A\right) \right. \nonumber  \\[2mm]
&&\mbox{\hspace{-15mm}} + \left. ~A \sqrt{\frac{2 }{\pi \delta \bar
v}}
\sum_{n=0}^{\infty}\frac{(t/\delta)_{-n-1/2}(1/2)_n}{(1)_{2n+1}}\left(
-\frac{2A^2}{\bar v\delta}\right)^{\!\!
n}\!~_{1\!}F_1\left(n+\frac{1}{2}, 2n + 2, A\right)\right. \nonumber  \\[2mm]
&&\mbox{\hspace{-15mm}} + \left. \frac{1}{\cos(\pi
t/\delta)}\left(\frac{2A^2}{\delta \bar v }\right)^{\!\! t/\delta}
\sum_{n=0}^{\infty}\frac{(t/\delta)_n}{(1)_{2n +  2t/\delta} (1)_n}
\left(\frac{2A^2}{\bar v\delta} \right)^{\!\!n}\!~_{1\!}F_1\left(n+
\frac{t}{\delta}, 2n + 1 + 2\frac{t}{\delta}, A\right) \right]\! .
\label{A13}
\\[-2mm] \nonumber\end{eqnarray}
In Appendix~A we also show that for small $\delta$, or large $t$, $\Phi^{(-)}_v $ approaches
the cumulative normal  distribution $\Phi(y_{\bar v}^-)$. The asymptotic behaviors ensure
us that the $\delta\rightarrow 0_+$ limit leads back to the original Black-Scholes
formula, as it should.

Another interesting situation arises for small $A$, where $S(t_a)
\approx E e^{-r_W (t_b - t_a)}$. Using the fact that
$\!~_{1\!}F_1(a,b,0)=1$, and that we may neglect for small $A$ the
last two sums in Eq.(\ref{IV9c}) and Eq.(\ref{A13}), we obtain that
both $\Phi^{(+)}_v$ and $\Phi^{(-)}_v$ approach the cumulative
normal distributions $\Phi(y_{\bar v}^+ \rangle)|_{A=0}$ and
$\Phi(y_{\bar v}^-)|_{A=0}$, respectively, implying that
$O_{\delta}(x_a,t_a)\rightarrow O^{(BS)}(x_a,t_a)|_{A=0}$. Options
with $A=0$ are known as {\em at-the-money-forward} options, i.e.
options whose strike price is equal to the current, prevailing price
in the underlying forward market. Inasmuch, whenever the option is
at-the-money-forward we regain back the formula of Black and
Scholes. Since many real transactions in the over-the-counter
markets are quoted and executed at or near
at-the-money-forward~\cite{McMillan02}, this explains some of the
empirical support for the Black-Scholes model .

In this connection it is interesting to note that for {\em put}
options, where the terminal condition is
\begin{eqnarray}
O(x_b,t_b) ~= ~\Theta(E - S(t_b))(E-S(t_b))\, ,
\label{IV9eee}
\\[-3mm]\nonumber\end{eqnarray}
we would have obtained the pricing equation in the form
\begin{eqnarray}\nonumber \\[-3mm]
O_{\delta}(x_a,t_a) ~= ~E e^{-r_W (t_b - t_a)} (1 - \Phi^{(-)}_v) - S(t_a)(1 -
\Phi^{(+)}_v)\, .
\label{IV9fff}
\\[-3mm]\nonumber\end{eqnarray}
This shows that our option-pricing model fulfills important
consistency condition known as the {\em put}-{\em call} parity
relation~\cite{B-P}
\begin{eqnarray}\nonumber \\[-3mm]
O^P_{\delta}\!(x_a,t_a) ~= ~O^C_{\delta}\!(x_a,t_a) - S(t_a) + E e^{-r_W (t_b -
t_a)}\, ,
\label{IV9ggg}
\\[-2mm] \nonumber\end{eqnarray}
where $O^P_{\delta}$ and $O^C_{\delta}$ denote put and call options,
respectively. Consequently, the case when $A \approx 0$ can be
equally phased as a situation with $O^C_{\delta}(x_a,t_a) \approx
O^P_{\delta}(x_a,t_a)$. We shall further see in the following section  that
the expansion factor $A/\sqrt{\bar v\delta }$ is
directly related to the so-called {\em moneyness} of the options.

Moneyness is a measure of the degree to which an option is likely to
have a nonzero value at the expiration date. In the Black-Scholes
formula, the moneyness (measured in the units of standard deviation)
is defined by~\cite{B-P}
\begin{eqnarray}\nonumber \\[-3mm]
m_{\rm BS} ~= ~\frac{y_v^+ ~+ ~y_v^-}{2} ~= ~\frac{\log[S(t_a)/E]~ +
~r_W t}{\sigma \sqrt{t}} ~= ~\frac{A}{ \sqrt{vt}}\, .
\label{cum26}
\\[-2mm] \nonumber \end{eqnarray}
For at-the-money-forward options, the moneyness is zero.
A positive value of $m_{\rm BS}$ corresponds to in-the-money-forward options
(i.e., options with positive monetary value) while a negative value represents
out-of-the-money options.

In our case $v=\sigma^2$ is a random variable and for given
$t$ it is distributed according to the Gamma distribution
 $f_{t\mu,t/\delta}(v)$. We may easily calculate the momenta of $|m_{\rm BS}|$.
The odd moments are
\begin{eqnarray}\nonumber \\[-3mm]
\frac{\langle |m_{\rm BS}|^{2n +1}\rangle}{(2n+1)!} ~= \
\frac{A^{2n+1}}{t^{n +1/2}}~\frac{{\langle}
{\sigma^{-2n-1}}{\rangle}}{(2n+1)!} ~= ~\left( \frac{A^2}{\delta
\bar v} \right)^{\!\! n} \sqrt{\frac{A^2}{\delta
\bar v}}~\frac{(t/\delta)_{-n-1/2}}{(1)_{2n+1}}\,
,
\\[-3mm] \nonumber \end{eqnarray}
so that the second sum in (\ref{IV9c}) and (\ref{A13}) corresponds
to an expansion in the odd momenta of the moneyness.
By taking into account that $(t/\delta)_{-n} = (t/\delta)_{n}/(t/\delta
- n)_{2n}$, the third sum in (\ref{IV9c}) and (\ref{A13})
is seen to be an expansion in the even momenta of
$m_{\rm BS}$.
%
%
Hence we can roughly characterize the three sums in the expansion
(\ref{IV9c}) and (\ref{A13}) as follows: the first sum is an
expansion involving the properties of the stochastic process
such as volatility, characteristic time, spread parameter  $\delta =q-1$),
while the remaining two expansions are expansions involving
the option contract characteristics (i.e., expiration time, strike price).

The computations presented above assume a single-pole structure of
(\ref{@CONT}), i.e. that $t/\delta$ is not half-integered. If
$t/\delta$ is half-integered, the  perturbation expansion
(\ref{IV9c}) clearly fails. The half-integered $t/\delta$ case must
be thus treated separately. For completeness, we state the
corresponding result for $\Phi^{(+)}_v$ coming from the double poles
in the $s$-plane of the integral (\ref{@CONT}). First of all, the
structure of $h^{(+)}_1$ stays the same because there are no
multi-poles present. The corresponding computation of $h^{(+)}_2$ is
not much more complicated than for simple-poles. Assuming that
$t/\delta = 1/2 + l, \; l \in {\mathbb{N}}$, the analysis reveals
that
\begin{eqnarray}\nonumber \\[-3mm]
&&\mbox{\hspace{-9mm}}\Phi^{(+)}_v = \frac{1}{2}\left[1 ~+
~\sqrt{\frac{\bar v \delta }{2\pi}} \sum_{n=0}^{\infty}\frac{(1/2 +
l)_{n+1/2}(1/2)_n}{(1)_{2n+1}}\left( -\frac{\bar
v\delta}{2}\right)^{\!\!
n}\!\!~_{1\!}F_1\!\left(n+\frac{1}{2}, 2n + 2, -A\right) \right. \nonumber \\[2mm]
&&\mbox{\hspace{-9mm}} + \left. ~2\sqrt{\frac{A^2}{2\pi \bar{v}
\delta}}
\sum_{n=0}^{l-1}\frac{(1/2+l)_{-n-1/2}(1/2)_n}{(1)_{2n+1}}\left(
-\frac{2A^2}{\bar v\delta}\right)^{\!\!
n}\!\!~_{1\!}F_1\!\left(n+\frac{1}{2}, 2n + 2, -A\right)\right. \nonumber  \\[2mm]
&&\mbox{\hspace{-9mm}}  + \left. \frac{(-1)^l}{\Gamma(l+1/2)}
\sqrt{\frac{A^2}{2\pi \bar v\delta}} \sum_{n=l}^{\infty}
\frac{\left[ \psi(n-l + 1) + \psi(n+1)
\right]}{\Gamma(n)\Gamma(n+l-1)}\left( \frac{A^2}{\bar
v\delta}\right)^{\!\! n}\! \!~_{1\!}F_1\!\left(n+\frac{1}{2}, 2n +
2, -A\right)\! \right]\! ,
 \label{IV9e}
\\[-1mm] \nonumber \end{eqnarray}
where $\psi(z) = \Gamma'(z)/\Gamma(z)$ is the Digamma
function~\cite{G-R}. It is possible to check numerically that for
$t/\delta \rightarrow 1/2 + l, \; l \in {\mathbb{N}}$ the relation
(\ref{IV9c}) equals to (\ref{IV9e}) (which is not true perturbatively!).
Thus $\Phi^{(+)}_v$ is a smooth and non-singular function at aforementioned critical times.

\section{Characteristic time} \label{SEc8}

When confronted with a practical option price problem, one
must decide whether or not $t$ (i.e, the time to expiration) is
large enough to deal satisfactorily with Gaussian
distributions. This is done by estimating the characteristic (or
crossover) time $t^*$, below which the Black-Scholes formula is
inapplicable and our solution becomes relevant. The estimate can be
done with the help of a Chebyshev expansion~\cite{Gnedenko}. We
define a rescaled time-compounded variable
\begin{eqnarray}
z ~= \frac{\Delta x(t) - r_{x_W}t}{\sqrt{t v
}}\, .
\label{IV9g}
\\[-3mm] \nonumber \end{eqnarray}
Removing the trivial drift we find  the difference between cumulative
distribution at time $t$ and that of the asymptotic Gaussian distribution
as an expansion~\cite{Gnedenko}
\begin{eqnarray}\nonumber \\[-3mm]
\Delta P_{\delta}(u) ~&\equiv& ~ \int_u^{\infty} dz \left[P_{\delta}(z,t) - \frac{1}{\sqrt{2\pi} }
e^{-z^2/2}\right]\nonumber \\[2mm]
&=& ~\frac{1}{\sqrt{2\pi} } e^{-u^2/2}
\left[\frac{Q_1(u)}{\sqrt{t}}  +  \frac{Q_2(u)}{t}  + \cdots +
\frac{Q_j(u)}{t^{j/2}} + \cdots \right]\, . \label{IV9gg}
\\[-3mm] \nonumber \end{eqnarray}
Here we have used the abbreviation $P_{\delta}(z,t) \equiv P_{\delta}(\Delta x = z,
\Delta t = t)$, with the time $t$ measured in the basic time units
$\Delta t_0$. Functions $Q_j$ are Chebyshev-Hermite polynomials, the coefficients of
which depend only on the first $j+2$ momenta of the random variable
$z$ appearing in the
elementary distribution
$P_{\delta}(z, \Delta t_0)$.
The first two expansion functions are
\begin{eqnarray}\nonumber \\[-3mm]
&&Q_1(u) ~= ~\frac{\kappa_3}{6}(1 - u^2)\, ,\nonumber \\[2mm]
&&Q_2(u) ~= ~\frac{10\kappa_3^2}{6!} ~u^5 ~+ \
\frac{1}{8}\left(\frac{\kappa_4}{3} - \frac{10\kappa_3^2}{9} \right)
u^3 ~+ ~\left(\frac{5\kappa^2_3}{24} - \frac{\kappa_4}{8} \right)
u\, .
\label{IV9k}
\\[-3mm] \nonumber \end{eqnarray}
Here $\kappa_3$ and $\kappa_4$ are {\em skewness} and {\em kurtosis}
of $P_{\delta}(z, \Delta t_0)$. Since the drift was removed, $P_{\delta}(z, \Delta t_0)$ is
symmetric and the skewness $\kappa_3 $ vanishes.

The characteristic time $t^*$ is now defined~\cite{B-P} as a time at
which the relative difference $|\Delta P_{\delta}(u)|/\Phi(-u)$
starts to be substantially smaller than $1$ (to be specific we
choose $1\%$), if $u$ is taken to be a typical endpoint of the
Gaussian central region, which we take as $u =1$. Since $|\Delta
P_{\delta}(u)|/\Phi(-u)   $ has  the value $\kappa_4/(t~\! 24 )$ for
$u= 1$, we identify $t^*$ with $\kappa_4$. This implies that for $t
\gg t^* = \kappa_4$ the Gaussian approximation is almost exact while
for $t < t^*=\kappa_4$ the Black-Scholes analysis is unreliable and
our new option pricing formula applies.


 To find $\kappa_4$ we calculate the moment generating
function  $G(p)$ associated with $P_{\delta}(z,\Delta t_0)$ via the Fourier transform
%
\begin{eqnarray}\nonumber \\[-1mm]
G(p) \ = \ \int_{-\infty}^{\infty} \d y \ e^{ipy} \ P(y,1)\ = \
\frac{2^{1/\delta} (\mu)^{1/2}}{(p^2 + 2\mu)^{1/\delta}}\, .
\label{IV12}
\\[-2mm]
\nonumber\end{eqnarray}
The cumulant generating function $W(p)$ is then the logarithm of
$G(p)$. The cumulants $c_n$ are obtained from the derivativesand of $W(p):$
\begin{eqnarray}\nonumber \\[-1mm]
c_n \ = \ \left.(-i)^n\frac{\d W(p)}{\d p^n}\right|_{p = 0} \;\;\;
{\Rightarrow} \;\;\;\;  &&c_2 \ = \ \frac{1}{\delta \mu} \ = \
\bar{v}\,  ,\;\; \;\;c_{2n} \ = \ \frac{(2n)!}{n
2^n}\frac{1}{\delta \mu^n} \ = \ \frac{(2n)!}{n 2^n} \ \!\bar{v}^{\! n} \delta^{n-1}\, ,
\nonumber \\[2mm] &&c_{2n+1} \
= \ 0,\; \;\; n \in {\mathbb{N}}\, . \label{cum22}
\\[-2mm]
\nonumber
\end{eqnarray}
The excess kurtosis has the value $\kappa_4 \equiv c_4/(c_2)^2 = 3\delta$,
implying that $t^*$ is equal to three times the spread of the distribution
$\delta=q-1$. This means that for large spread one can apply the Black-Scholes formula
only a long time before expiration (long maturity options).

It is worth noting that the present analysis is not applicable in
cases when the distribution of returns is of the Tsallis type. This
has heavy power-like tails, so that the second moment may be
infinite and the above Chebyshev's expansion may not exist. In the
present case the Tsallis distribution is in Fourier space and all
cumulants are finite leading to the above estimate of $t^*$.

 Our cumulant calculation (\ref{cum22}) also further clarifies the
meaning of the expansions (\ref{IV9c}) and (\ref{A13}). In fact, the
first sum in both (\ref{IV9c}) and (\ref{A13}) corresponds directly
to the expansion in cumulants $c_{2n}, ~n \in {\mathbb{N}}$. This
is because
\begin{eqnarray}\nonumber \\[-3mm]
\frac{\delta ~\! c_{2n}}{\Gamma(n)2^{2n}} ~= ~(1/2)_{ n}
\left(\frac{\bar v \delta}{2} \right)^{\! n} \,
.\label{cum23}
\\[-3mm] \nonumber \end{eqnarray}
Since $\bar v = \beta$, we can alternatively
view the aforementioned expansion as an expansion in small
$\beta$, i.e., as a ``high-Tsallis-temperature"-expansion. This establishes
contact with the market temperature introduced in Ref.~\cite{TIMX}.

Let us finally mention that the characteristic time can be crudely
but fairly rapidly estimated from the shape of the variance (or
volatility) distribution. In particular, the relative width of the
distribution $f_{t\mu, t/\delta}(v)$ at time $t$ is
\begin{eqnarray}\nonumber \\[-3mm]
\frac{\overline{(v - \overline{v})^2}}{\overline{v}^2} \ = \
\frac{\delta}{t}\, . \label{chartime2}
\\[-3mm] \nonumber \end{eqnarray}
If the LHS of (\ref{chartime2}) is much smaller than $1$ then the
distribution is effectively $\delta$-function and the volatility is
a constant as assumed in the Black-Scholes analysis. However, when
the relative width starts to be of order $1$ the Black-Scholes model
ceases to be valid. This happens at the time $t^* \approx \delta$
which is consistent with our previous estimate.

\section{Comparison with empirical market data} \label{SEc6b}

\begin{figure}[h]
\hspace{-0.2cm}\centerline{\resizebox{16.7cm}{7.1cm}{\includegraphics{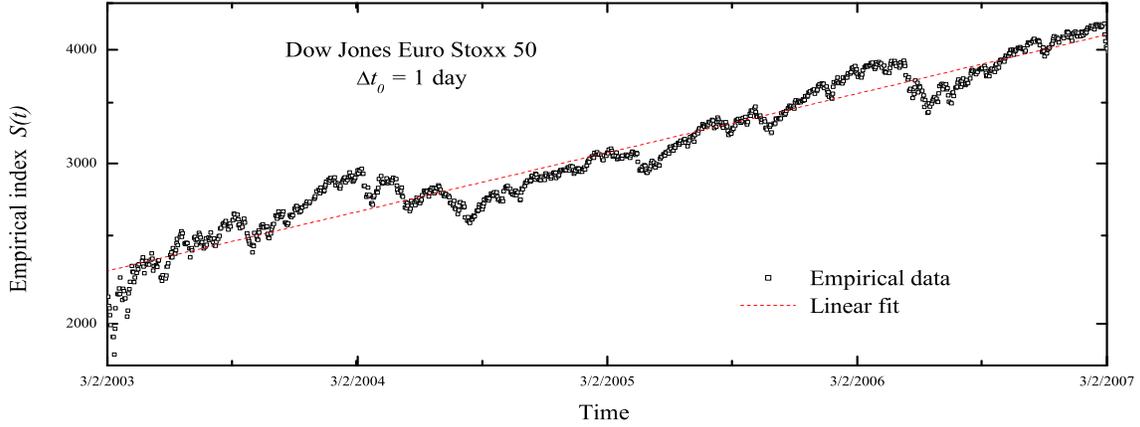}}}
\vspace{-0.9cm}\caption{\label{DJI} Logarithmic plot of Dow Jones
Euro Stoxx 50 index $S(t)$ over
 $4$ years (5 Feb 2003 - 21 March 2007, in total $1057~$trading
days) with sampling interval $\Delta t_0 = 1$ day. The index is
fitted by a straight line, implying an exponential growth at an
annual rate $\approx 20\%$.} \vspace{0.2cm} \hrule
\end{figure}

\begin{figure}[h]
\hspace{-0.2cm}\centerline{\resizebox{16.7cm}{7.2cm}{\includegraphics{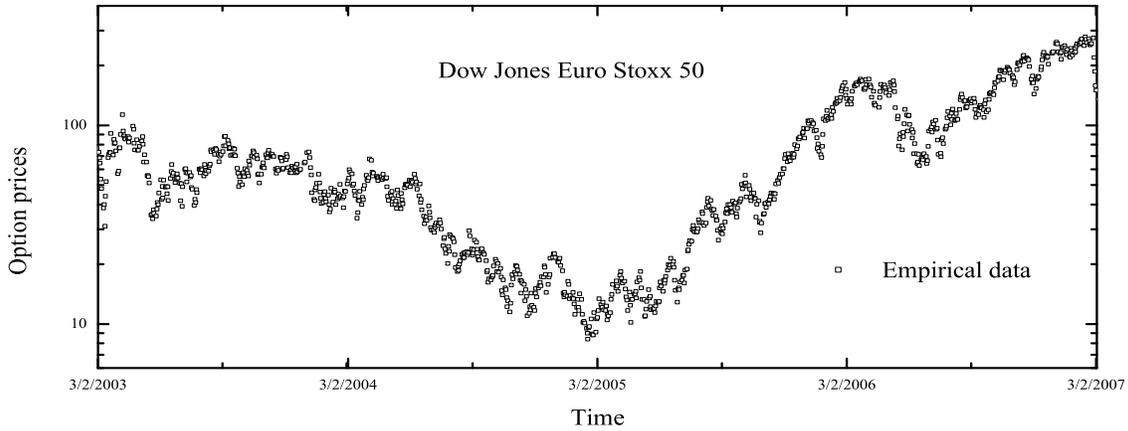}}}
\vspace{-0.9cm}\caption{\label{DJOP} Option prices for an European
option with a strike price $4200~$EUR and expiry on 21 December
2007. The underlying of the option is the Dow Jones Euro Stoxx 50
with the time series given in Fig.~\ref{DJI}.} \vspace{0.2cm} \hrule
\end{figure}

It is interesting to compare our solution (\ref{IV8}) for European
call options with realistic market data. Consider the
option prices for an European option whose underlying is the
Dow Jones Euro Stoxx 50 with the time series shown in
Fig.~\ref{DJI}.
The associated empirical option prices are plotted
in Fig.~\ref{DJOP}. It is clear that because of the noise in the data
no option pricing formula will fit the market prices
perfectly.
Even for a very good fit the option price would not be
the most meaningful measure for the quality of the pricing model. Instead,
one should test directly the hedging qualities of the pricing model~\cite{B-P}.
To this end one introduces the $\Delta$-hedge
\begin{eqnarray}\nonumber \\[-3mm]
\Delta(\tau) \ \equiv \ \frac{\partial O_{\delta}(\tau)}{\partial
S(\tau)} \, .\label{delta-hedge}
\\[-3mm] \nonumber \end{eqnarray}
%
Given $ \Delta (\tau )$ we construct a so-called $ \Delta $-hedged
portfolio   $\Pi(\tau )$ by  mixing  stocks and options so that
%
%
\begin{eqnarray}\nonumber \\[-3mm]
\Pi(\tau) = O_{\delta}(\tau) - \Delta(\tau)S(\tau)\, .
\label{portfolioII}
\\[-3mm] \nonumber \end{eqnarray}
The cash amount $\Pi$ can be interpreted as the value of a portfolio of a trader who at time $\tau$
bought one option $O_{\delta}(\tau)$ and sold an amount $\Delta(\tau)$ of the
underlying with price $S(\tau)$. In the ideal situation, i.e., neglecting the time delays
in the determination of $ \Delta (\tau )$ and the adaptation of
the portfolio, and ignoring the transaction costs, this portfolio
is perfectly hedged, i.e., it is free of fluctuations since
the fluctuations of $S$ and $O_{\delta}$ cancel each other~\cite{B-P}.
The growth of the portfolio is therefore deterministic and proceeds at the
riskfree rate $r_W$:
\begin{eqnarray}\nonumber \\[-3mm]
\frac{\d \Pi(\tau)}{\d \tau} \ = \ r_{W} \ \! \Pi(\tau)\, .
\\[-3mm] \nonumber \end{eqnarray}
Any other growth rate would yield  arbitrage possibilities.
The amount $\Delta(\tau)$ deduced directly from empirical data
should thus yield portfolio $\Pi(\tau)$ that is almost precisely
$e^{r_{W} \tau }$ (modulo multiplicative pre-factor). Consequently, $\Delta(\tau)$ computed from a
{\em good} option pricing formula must give
$\Pi(\tau)$ that is also close to the $e^{r_{W} \tau }$ behavior. In our case the $\Delta$-hedge
is
\begin{eqnarray}\nonumber \\[-3mm]
\Delta \ &=& \ \Phi^{(+)}_v(A)  \ + \ \left[S \ \frac{\partial
 \Phi^{(+)}_v(A)}{\partial A} \ -
\ E e^{-r_W t} \ \frac{\partial \Phi^{(-)}_v(A)}{\partial A}
 \right] \frac{\d A}{\d S} \
= \ \Phi^{(+)}_v(A) \, ,
\\[-3mm] \nonumber \end{eqnarray}
since the expression in the square bracket vanishes due to
Eqs.~(\ref{phi-plus}), (\ref{phi-minus}) and the identities [see
e.g., Ref.~\cite{G-R}]
\begin{eqnarray}\nonumber \\[-3mm]
&&\frac{\partial \!~_{1\!}F_1(a, b,
A)}{\partial A} \ = \ \frac{a}{b} \!~_{1\!}F_1(a + 1, b + 1,
A)\, , \\[2mm]
&&\!~_{1\!}F_1(a, b,
A) \ = \ e^{A} \!~_{1\!}F_1(b-a, b,
-A) \ = \  \frac{S}{E} \ \!e^{r_W t} \!~_{1\!}F_1(b-a, b,
-A)\, , \\[2mm]
&&a \ \!\!~_{1\!}F_1(a + 1, b,
A)\ = \ a \ \!\!~_{1\!}F_1(a, b,
A) \ + \ A \ \!\frac{\partial\!~_{1\!}F_1(a, b,
A)}{\partial A} \  \, .
\\[-3mm] \nonumber \end{eqnarray}
%

\begin{figure}[h]
\hspace{-0.2cm}\centerline{\resizebox{17cm}{7.4cm}{\includegraphics{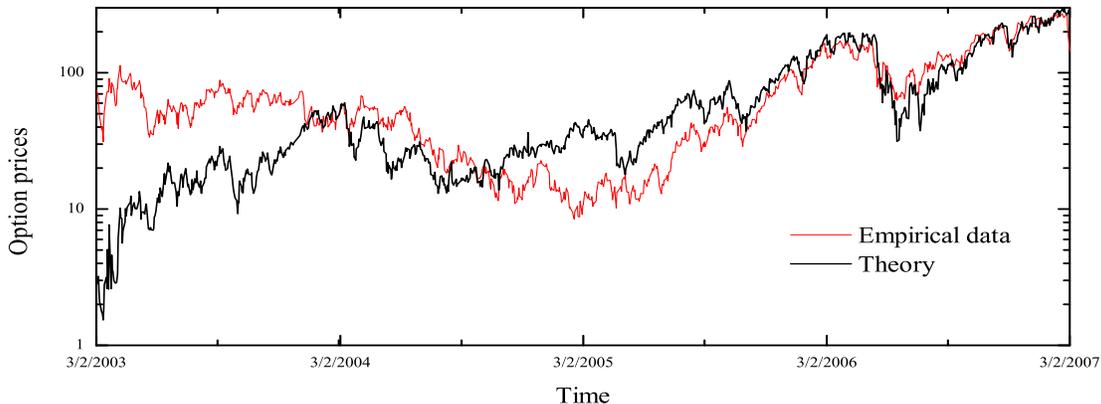}}}
\vspace{-1.0cm}\caption{\label{option-pricing-fit} The best
$O_{\delta}$ fit for the Dow Jones Euro Stoxx 50 option prices (cf.
Fig.~\ref{DJOP}). On this scale the Black-Scholes solution basically
coincides with the $O_{\delta}$ prediction.} \vspace{0.2cm} \hrule
\end{figure}

 Let us note that by comparing (\ref{portfolioII}) with the
option-pricing formula (\ref{IV8}) we can write the portfolio at the
time $\tau$ in the explicit form
\begin{eqnarray}\nonumber \\[-3mm]
\Pi(\tau ) \ = \ - E e^{-r_W(t_b - \tau)} \Phi^{(-)}_v(\tau)\, .
\label{portfolio}
\\[-3mm] \nonumber \end{eqnarray}
Imperfectness of the hedging induced by the pricing formula (\ref{IV8})
thus depends on the actual
behavior of $\Phi^{(-)}_v(\tau)$ in time. In the Black-Scholes model
the corresponding $\Phi(y^-_v)$ is effectively $\tau$-independent
due to assumed form of the geometric Brownian motion for $S(\tau)$. In our model
the situation is less obvious because the additional stochastic
process due to volatility may substantially spoil the time
independence. To see how big is the amount of the residual risk
implied by our pricing model we turn now back to our empirical data.
The best option-pricing solution $O_{\delta}$ fit for the empirical
option-price data from Fig.~\ref{DJOP} is depicted in
Fig.~\ref{option-pricing-fit}. In the plot Fig.~\ref{option-pricing-fit} (and the plots to follow)
we have used $O_{\delta}$ calculated to $25$th perturbation order with the double-poles removed.
The obtained result is quite robust
with the residual error smaller that $0.05\%$.

\begin{figure}[h]
\centerline{\resizebox{17cm}{7.4cm}{\includegraphics{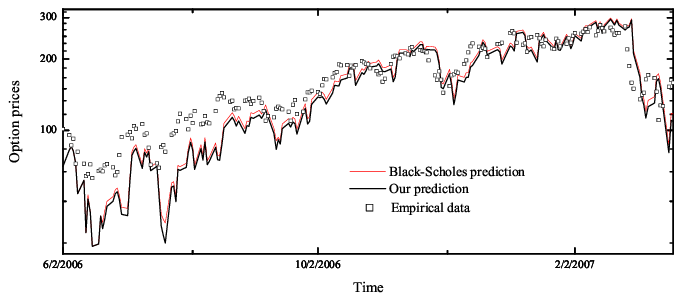}}}
\vspace{-1.6cm}
\centerline{\!\!\!\!\!\!\!\!\!\!\!\!\!\!\!\!\!\!\!\!\!
\resizebox{19.2cm}{7.5cm}{\includegraphics{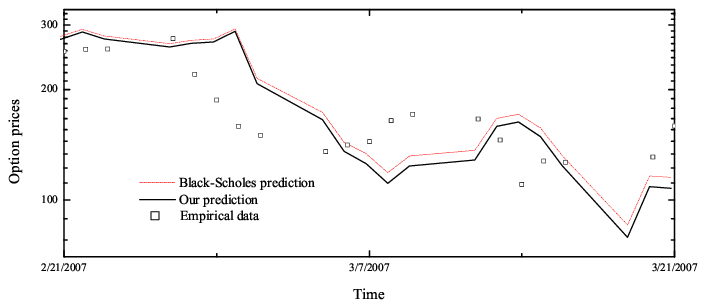}}}
\vspace{-12mm} \caption{\label{option-pricing-fit-detail} Two
successive details of the departure of the $O_{\delta}$ fit from the
Black-Scholes values. Black-Scholes option prices are taken with $v
\equiv \overline{v}$. The roughly $5\%$ departure is clearly visible
already $9-10$~months before the expiry date.} \vspace{0.2cm} \hrule
\end{figure}

Using as inputs: $E = 4200~$EUR, $r_W = 4.5$\%/year and $t_b =
1270~$trading days ($t_a = 5~$Feb$~2003 \equiv 0$), the
corresponding free parameters are then best fitted with $\delta =
69.43$, $\mu = 351.29$ (i.e., $\bar{v} = 0.000041$). The fit can be
further optimized when newly arrived option pricing data are taken
into account. Details of the departure of the $O_{\delta}$
prediction from the Black-Scholes fit is depicted in
Figs.~\ref{option-pricing-fit-detail}. 
The goodness of the fit can
be estimated, for instance, by the method of least squares. The
correspondent likelihood function $\chi^2$ is for the $O_{\delta}$
fit in the period $21~$Dec$~006 - 21~$Jan$~2007$ smaller by $1.2\%$
in comparison with the Black-Scholes best fit. In the period
$21~$Jan$~2007 - 21~$Feb$~2007$ the difference in $\chi^2$ is
$1.8\%$, and in the period $21~$Feb$~2007 -21~$Mar$~2007$ it is
already $3.3\%$. This trend seems to get even more pronounced for
periods closer to the expiry date.

 According to Section~\ref{SEc8}, the characteristic time corresponds
to the time scale where non-Gaussian effects begin to smear out and
beyond which the CLT begins to operate. For the call options at hand
$t^* = 3\delta \sim 210$~days, which,  is particularly large and
although the actual value will get further adjust with newly arrived
option data, Fig.~\ref{option-pricing-fit-detail} indicates that
$t^*$ will not get dramatically changed. Note also, that despite the
fact that $t^* = 210$~days, the departure effect is visible already
$10$ months before the maturity.

\begin{figure}[h]
\hspace{-0.2cm}\centerline{\resizebox{17cm}{7.4cm}{\includegraphics{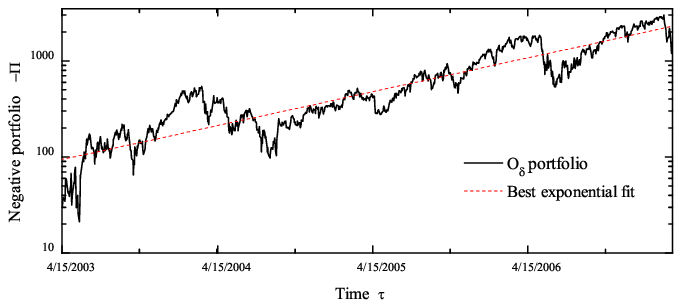}}}
\vspace{-1.0cm}\caption{\label{delta-hedging} Negative daily
portfolio $-\Pi$ constructed from the $O_{\delta}$ pricing formula
(\ref{IV8}) through $\Delta$-hedging prescription
(\ref{delta-hedge}) and (\ref{portfolioII}). The best exponential
fit confirms the annual interest rate $4.5\%$. The relative
fluctuations $\delta r_W(\tau)/r_W = 0.007 \pm 0.002$.}
\vspace{0.2cm} \hrule
\end{figure}

 By having the best $O_{\delta}$ fit we can construct the
daily portfolio $\Pi$ according the prescription
(\ref{portfolioII}). The resulting portfolio is shown in
Fig.~\ref{delta-hedging}. 
To quantify the fluctuations it is
convenient to write $\Pi(\tau) = - E\Phi^{(-)}_v(\tau)\ \! e^{-(t_b
- \tau)r_W} = - E\Phi(y^-_v)\ \! e^{-(t_b - \tau)(r_W + \delta
r_W(\tau))}$. The relative interest-rate fluctuations $\delta
r_W/r_W$  implied by the stochastic nature of the volatility are
according to Fig.~\ref{delta-hedging}, $0.007(\pm 0.002)$.
Consequently we can conclude that for the data at hand the
$O_{\delta}$ pricing formula induces a $\Delta$-hedging strategy
that is close to being optimal.

%

\section{Conclusions and Outlook} \label{SEc7}

We have developed the theory of an option pricing model with a
stochastic volatility following a Chi distribution. The corresponding volatility
variance is then distributed with an ubiquitous Gamma distribution. Our direct
motivation was drawn from the high-frequency S\&P 500 volatility fluctuation data,
whose distribution is well fitted in this way. There are two
interesting implications resulting from such volatility
fluctuations. First, the returns of the corresponding asset prices
exhibit semi-heavy tails around the peak (i.e., leptocurtic
behavior) which mimics empirically observed long-range correlations.
At the same time, the returns preserve some of typical features of
the original Black-Scholes model, namely they follow a linear
It\={o} stochastic equation with multiplicative noise (though with
stochastic volatility) and a continuous stock dynamics. Second, the
associated density operator in momentum space is of the Tsallis
type. The inverse Tsallis temperature then agrees with the average
variance of the distribution.

 Our main result is a generalized Black-Scholes pricing
formula that takes into account the above volatility behavior. With
the help of a Mellin transform we were able to find an
ensuing analytic solution for the price of European call options.
The result is expressed as a series in the  higher normalized
cumulants and in higher moments of moneyness. Due to a
spread parameter $\delta$ related to the extra kurtosis of the log-return data,
our model is capable of calibration to a richer set of observed market histories than the
simple Black-Scholes model, which is a special case corresponding to
a zero spread $\delta $, or to a very long time horizon ($t \gg t^*\sim
\delta$).
Comparisons with other time-dependent volatility models such as
ARCH~\cite{Engle82}, GARCH~\cite{Bollerslev86} or multiscale
GARCH~\cite{Z-L03} will be addressed in future work.

 In the light of recent works on superstatistics we should mention
that the density operator representation (\ref{3.3}) reveals the
superstatistic nature of the Tsallis distribution in momentum space.
Recently, C.~Beck~\cite{1,2} and C.~Beck and E.~Cohen~\cite{3},
prompted by the works by G.~Wilk and Z.~Wlodarczyk~\cite{4}, have
suggested that the origin of certain heavy-tail distributions should
be understood as weighted averages of the usual exponential
statistics. Such averages have been used also in the textbook
\cite{HK} to calculate option prices for non-Gaussian distributions
\cite{REPI}. Recently, F.~Sattin~\cite{5} rephrased the same
procedure in terms of evolving systems embedded within a static but
non-trivial background. All these approaches commonly strive to
interpret broad distributions as a result of an averaging of usual
exponential (Gibbs) distributions over certain fluctuating (random)
parameter. This is the procedure running under the name
superstatistics~\cite{3,5}. Although frequently used,
superstatistics procedure is still far from being systematized and,
in fact, smearing distributions are usually chosen to fit
experimental data. The theory of option pricing in this work may
therefore be viewed as an example for superstatistics with a smeared
Gaussian representing the probability distribution of volatilities.

 Finally, the present scenario also provides an
interesting meaning to the Tsallis escort parameter $q$. In option
pricing models where stock fluctuations are directly fitted by
a Tsallis distribution of returns~\cite{HK,Borland}, the $q$
parameter is a fit  parameter with a typical value around $1.5$.
In the present case where the momentum distribution is of the Tsallis
type, $q-1$ is proportional to the characteristic time $t^*$ below which
the Gaussian treatment of stock fluctuations is inadequate.

\begin{acknowledgments}
One of us (P.J.) acknowledges discussions with Prof.~T.~Arimitsu and
Dr.~X.~Salier, and financial supports from the Doppler Institute in
Prague, from the Ministry of Education of the Czech Republic
(research plan no. MSM 6840770039), and from the Deutsche
Forschungsgemeinschaft under grant Kl256/47. We all wish to thank
Dr.~X.~Salier and Dr.~A.~Garas for providing us with the data.
\end{acknowledgments}

\appendix

\section{Fokker-Planck equation for stochastic
Eq.(1) and some consequences}\label{Ap1}

Some remarks may be useful concerning the stochastic equations for volatility and
variance stated in the introduction.

 Let us assume that the variance $v$ is driven by the It\={o} stochastic process
\begin{eqnarray}\nonumber \\[-4mm]
\d v(t) \ = \ \gamma(t)[\nu(t) - \mu(t) v(t) - a(v(t), \mu(t),
\nu(t))] \d t \ + \ \sqrt{2 \gamma(t) v(t)} \; \! \d W(t)\, .
\label{AA00}
\\[-3mm] \nonumber\end{eqnarray}
Here $\gamma(t), \mu(t)$ and $\nu(t)$  are arbitrary non-singular
positive real functions on $\mathbb{{R}}^+$. The function $a(v(t), \mu(t),
\nu(t))$ is
assumed to be a non-singular function of its arguments.
The corresponding Fokker-Planck equation
for the distribution function $\rho(v,t)$ reads
\begin{eqnarray}\nonumber \\[-4mm]
\frac{\partial \rho(v,t)}{\partial t} \ = \ \frac{\partial}{\partial
v}\left\{ \gamma(t)\left[v \mu(t) - \nu(t) + a(v, \mu(t),
\nu(t))\right] \,\! \rho(v,t)\right\}  \ + \ \frac{\partial^2
}{\partial v^2}\left[\gamma(t) v \,\! \rho(v, t) \right]\, .
\label{AA01}
\\[-3mm] \nonumber\end{eqnarray}
Let us choose the function $a(v, \mu(t), \nu(t))$ to be
\begin{eqnarray}\nonumber \\[-4mm]
a(v, \mu,\nu ) \ = \ v \frac{(\log \mu)'}{\gamma} \ &+& \ v
e^{v\mu}\left\{\left[\Gamma(\nu)\log\left(\frac{v\mu}{\nu}\right) +
\Gamma(\nu, v\mu)(\psi(\nu)
- \log(v \mu))\right](v\mu)^{-\nu}\right.  \nonumber \\[2mm]
&-& \left. \frac{ ~_2F_2(\nu,\nu;\nu+1,\nu+1; -v\mu)}{\nu^2}
\right\} \frac{(\nu)'}{\gamma} \, ,\label{AA02}
\\[-3mm] \nonumber\end{eqnarray}
where $\Gamma(x,y)$ is the incomplete gamma function, $\psi(x)$  the
digamma function, and $\!\!~_2F_2$  a hypergeometric function~\cite{G-R})
the solution of (\ref{AA01}) has the
form
\begin{eqnarray}\nonumber \\[-4mm]
\rho(v,t) \ =
 \frac{1}{\Gamma(\nu(t))} ~[\mu(t)]^{\nu(t)} v^{\nu(t)
-1} e^{-\mu(t) v}\,
.\label{AA03}
\\[-3mm] \nonumber\end{eqnarray}
This is precisely  the Gamma distribution $f_{\mu(t),\nu(t)}(v)$
in Eq.~(\ref{Gammad})
describing the temporal statistical evolution of the random variable
$v$. Note also that $\gamma(t)$ does not enter in the solution.

By applying the It\={o} formula of the stochastic calculus~\cite{BO1} one easily
obtains the corresponding It\={o} stochastic equation for
the volatility $\sigma = \sqrt{v}$ in the form
\begin{eqnarray}\nonumber \\[-4mm]
\d \sigma(t) \ = \
\frac{\gamma(t)}{2}\left[\frac{(\nu(t)-1/2)}{\sigma(t)} - \mu(t)
\sigma(t) - \frac{a(\sigma^2(t), \mu(t), \nu(t))}{\sigma(t)}\right]
\d t \ + \ \sqrt{\frac{\gamma(t)}{2} } \; \! \d W(t)\, ,
\label{AA04}
\\[-3mm] \nonumber\end{eqnarray}
from which we derive the Fokker-Planck equation for the
distribution of the volatility $\sigma= \sqrt{v} $:
\begin{eqnarray}\nonumber \\[-4mm]
\frac{\partial \rho(\sigma,t)}{\partial t} \ &=& \
\frac{\partial}{\partial \sigma}\left\{
\frac{\gamma(t)}{2}\left[\sigma \mu(t) - \frac{(\nu(t)
-1/2)}{\sigma} + \frac{a(\sigma^2, \mu(t),
\nu(t))}{\sigma}\right] \,\! \rho(\sigma,t)\right\}\nonumber \\[2mm]
&+& \ \frac{\partial^2 }{\partial \sigma^2}\left[\frac{\gamma(t)}{4}
\,\! \rho(\sigma, t) \right]. \label{AA05}
\\[-3mm] \nonumber\end{eqnarray}
Of course, this equation is simply related to (\ref{AA01}) by the substitution
$v = \sigma^2$. Equation Eq.(\ref{AA05}) has the solution
\begin{eqnarray}\nonumber \\[-4mm]
\rho(\sigma,t)\ =
 \ 2\sigma
f_{\mu(t),\nu(t)}(\sigma^2) \, , \label{AA06}
\\[-3mm] \nonumber\end{eqnarray}
which is also known as the Chi distribution~\cite{feller,wik}.

 The particular solutions (\ref{AA03}) and (\ref{AA06}) are
the only ones fulfilling the asymptotic conditions
$\rho(v,t)|_{t\rightarrow \infty} =  \delta(v- \bar{v})$ and
$\rho(\sigma,t)|_{t\rightarrow \infty} = \delta(\sigma -
\bar{\sigma})$, respectively. They are also both normalized to
unity. Note that the uniqueness of the solutions (\ref{AA03}) and
(\ref{AA06}) ensures that the stochastic equations (\ref{AA00}) and
(\ref{AA05}) have weak solutions.

 Anticipating the empirical form presented in Section~\ref{SEc2a}
we shall now assume that at large $t$ the functions $\mu(t)$
and $\nu(t)$ behave like $\mu(t) \approx \mu t$ and $\nu(t) \approx \nu t$.
Without loss of generality we also set $\gamma(t) = 1$. In this case
the large-$t$ behavior of $a(v, \mu(t), \nu(t))$ can be easily
found. Using the asymptotic formulas~\cite{Luke}
\begin{eqnarray}\nonumber \\[-4mm]
&&\psi(\nu t)|_{t\rightarrow\infty} \ \approx \ \log(\nu t)  \, ,
\nonumber
\\[2mm]
&&\!\!~_2F_2(\nu t,\nu t;\nu t+1,\nu t+1; -v\mu
t)|_{t\rightarrow\infty} \ \approx \ e^{-v\mu t}\left(1 \ + \
\frac{v \mu}{\nu}\right)^{\! 2}\, , \nonumber \\[2mm]
&&\Gamma(\nu t, v \mu t)|_{t\rightarrow\infty} \ \approx \ e^{-\nu
t}(\nu t)^{\nu t - 1/2} \sqrt{2\pi} \ - \ (\nu t)^{-1}(v \mu t)^{\nu
t} e^{-v \mu t}\left(1 \ + \ \frac{v \mu}{\nu}\right)
 \, , \label{AA07}
\\[-3mm] \nonumber\end{eqnarray}
we obtain that $a(v, \mu(t), \nu(t)) |_{t\rightarrow\infty}
 \ \approx \  {\mathcal{O}}(1/t) $.
In this limit, the stochastic differential
equation (\ref{AA00}) for the variance
 corresponds to the
Cox-Ingersoll-Ross process~\cite{CIR}. Although this asymptotic
equation resembles Heston's stochastic volatility
model~\cite{Heston1}, there is a difference:
in the drift term $v(t)$ and $\mu(t)$ are linear
functions of $t$ rather than constants.

Let us assume the same linear behavior $\mu(t) \approx \mu t$ and $\nu(t)
 \approx \nu t$ at small $t$ (more precisely for
 $\nu t \ll 1$). For convenience we set $\mu/\nu \equiv \overline{v}$, as in
Section~\ref{SEc3}. The small-$t$ behavior of $a(\ldots)$ can be
easily found from the asymptotic
formulas~\cite{Luke}
\begin{eqnarray}\nonumber \\[-4mm]
&&\psi(\nu t)|_{\nu t\rightarrow 0} \ \approx \ - \frac{1}{\nu t} +
C \, , \nonumber
\\[2mm]
&&\!\!~_2F_2(\nu t,\nu t;\nu t+1,\nu t+1; -v/\overline{v} \ \!\nu t)|_{\nu
t\rightarrow 0} \ \approx \ 1\, , \nonumber \\[2mm]
&&\Gamma(\nu t, v/\overline{v}\ \! \nu t)|_{\nu t \rightarrow 0} \ \approx \  - \mbox{log}[\nu t] - \mbox{log}[v/\overline{v} ] - C
 \, , \label{AA09}
\\[-3mm] \nonumber\end{eqnarray}
where $C = 0.57731...$ is the Euler-Mascheroni constant. Then
 $a(v, \mu(t), \nu(t)) |_{\nu t \rightarrow 0}
 \ \approx \  - v/(\overline{v}\ \! t \nu) + {\mathcal{O}}(1/\nu t)$.
This shows that in the small-$t$ limit, the drift term is
dominated by the function $a(v, \mu(t), \nu(t))$. Let us finally note that
the stochastic equation (\ref{AA04}) corresponds to an additive
process, rather than to a multiplicative one.

\section{Calculation of $\Phi^{(-)}_{v}$}

Let us calculate the function  $\Phi^{(-)}_v$in Eq.~(\ref{A13})
appearing in the generalized Black-Scholes formula (\ref{IV8}). The
procedure is analogous to that for $\Phi^{(+)}_v$. We start from
Eq.~(\ref{IV9}) which reads now
\begin{eqnarray}
\Phi^{(-)}_v ~&=&
 ~\frac{1}{2} ~+ ~{{\cal P}}\int_{-\infty}^{\infty}\frac{\d
p}{2\pi i} ~F^{(-)}_{t\mu, t/\delta}(p)
~ \frac{e^{-p^2/2}}{p} \, ,\label{A00}
\\[-3mm] \nonumber\end{eqnarray}
where
 \begin{eqnarray}
F^{(-)}_{t\mu, t/\delta}(p)\ &=& \ 2
\frac{\mu^{t/\delta}}{\Gamma(t/\delta)} \int_{0}^{\infty}\d \omega \
\omega^{2t/\delta -1} ~e^{-\mu \omega^2} \
\exp\left[i\frac{p}{2}\left(\frac{2A}{\omega} - \omega
\right)\right] \nonumber \\[2mm]  &\equiv& \ 2
\frac{\mu^{t/\delta}}{\Gamma(t/\delta)} ~g^{(-)}_\mu(p,A)~,\label{A11}
\\[-3mm] \nonumber\end{eqnarray}
To find $g^{(-)}_\mu(p,A)$ we first calculate the integral
 \begin{eqnarray}\nonumber \\[-3mm]
\int_{0}^{\infty}\d \mu ~\mu^{s-1}g^{(-)}_\mu(p,A) ~&=& ~\Gamma(s)
\int_{0}^{\infty}\d \omega~\omega^{2t/\delta -2s -1}\
\exp\left[i\frac{p}{2}\left(\frac{2A}{\omega} - \omega
\right)\right] \nonumber \\[2mm]
&=& ~\Gamma(s)(2A)^{t/\delta - s} \int_{-\infty}^{\infty}\d u \
e^{-ip\sqrt{2A}\cosh u - (2s - 2t/\delta)u}\nonumber \\[2mm]
&=& ~2\Gamma(s)(e^{-i\pi} 2A)^{t/\delta - s } K_{2s -
2t/\delta}(p\sqrt{2A}) \, ,  \;\;\;\;\;\;\; (p>0)\, , \label{A12}
\\[-3mm] \nonumber\end{eqnarray}
where $K_{ \alpha }(z)$ is the modified Bessel
function~\cite{Watson}.  The result (\ref{A12}) holds in a strip
$t/\delta - 1/2 < {\Re}(s) < t/\delta + 1/2$.
Similar calculations for $p<0$ yields
 \begin{eqnarray}\nonumber \\[-3mm]
\int_{0}^{\infty}\d \mu ~\mu^{s-1}g^{(-)}_\mu(p,A) ~&=& \
2\Gamma(s)(e^{i\pi} 2A)^{t/\delta - s } K_{2s -
2t/\delta}(-p\sqrt{2A}) \, ,  \;\;\;\;\;\;\; (p<0)\, .
\\[-3mm] \nonumber\end{eqnarray}
Again, the result is true in a strip $t/\delta - 1/2 < {\Re}(s) <
t/\delta + 1/2$. With the help of the Mellin inverse transform we
now find
 \begin{eqnarray}\nonumber \\[-3mm]
g^{(-)}_\mu(p,A)~= ~\int_{c-i\infty}^{c+i\infty}\frac{\d s}{2\pi i} \
2\mu^{-s}\Gamma(s)\!\left\{
          \begin{array}{l}
            (e^{-i\pi} 2A)^{t/\delta - s } K_{2s -
2t/\delta}(p\sqrt{2A})
 ,~~~p>0.\\[2mm]
(e^{i\pi} 2A)^{t/\delta - s } K_{2s - 2t/\delta}(-p\sqrt{2A})
 ,~~~p<0,
          \end{array}
       \right\}
\\[-3mm] \nonumber\end{eqnarray}
where $c \in (t/\delta - 1/2, t/\delta+ 1/2)$. Inserting this back into
(\ref{A11}) and (\ref{A00}) we obtain
 \begin{eqnarray}\nonumber \\[-3mm]
\Phi^{(-)}_v  &=&  \frac{1}{2} + \frac{2\mu^{t/\delta}}{\Gamma(t/\delta)}\
\!{{\cal P}}\!\!\int_{-\infty}^{\infty} \frac{\d p}{2\pi i}
\frac{e^{-p^2/2}}{p} \int_{c-i\infty}^{c + i\infty}\!\!\frac{\d
s}{\pi i}~\mu^{-s}\Gamma(s)\!\left\{
          \begin{array}{l}
            (e^{-i\pi} 2A)^{t/\delta - s } K_{2s -
2t/\delta}(p\sqrt{2A})\\[2mm]
(e^{i\pi} 2A)^{t/\delta - s } K_{2s - 2t/\delta}(-p\sqrt{2A})
          \end{array}
        \right\} \nonumber \\[2mm]
=&&  \!\!\!\!\!\!\!\frac{1}{2} + \frac{4}{\Gamma(t/\delta)}\int_{c-i\infty}^{c +
i\infty}\!\!\frac{\d s}{2\pi i}~\Gamma(s) (2A\mu)^{t/\delta
-s}\sin[\pi(s -t/\delta )] \int_{0_+}^{\infty} \frac{\d q}{\pi }
\frac{e^{-p^2/2}}{p}
 K_{2s - 2t/\delta}(p\sqrt{2A})\, .\nonumber \\
\\[-2mm] \nonumber\end{eqnarray}
The $p$-integration can be carried out explicitly yielding ($\zeta
\equiv s-t/\delta$)
 \begin{eqnarray}\nonumber \\[-3mm]
&&\mbox{\hspace{-1cm}}\sin(\pi \zeta)\int_{0_+}^{\infty} \d p
~\frac{e^{-p^2/2}}{p}\
 K_{2\zeta}(p\sqrt{2A})\nonumber \\[2mm] &&\mbox{\hspace{-1cm}}= \
\frac{\sin(\pi \zeta)}{4}
\left[A^{-\zeta}\Gamma(-\zeta)\Gamma(2\zeta) ~_{1\!}F_1(-\zeta, 1-
2\zeta, A) + A^{\zeta}\Gamma(\zeta)\Gamma(-2\zeta)
~_{1\!}F_1(\zeta, 1 + 2\zeta, A)\right]\nonumber \\[2mm]
&&\mbox{\hspace{-1cm}}= ~-\frac{\sqrt{\pi}}{8 \zeta }
\left[\!\left(\frac{A}{4}\right)^{\!\!\zeta}\Gamma\!\left(\mbox{
$\!\!\frac{1}{2}$} -\zeta\right)\!\!~_{1\!}F_1(\zeta, 1 + 2\zeta, A)
+ \left(\frac{A}{4} \right)^{\!\!-\zeta}\Gamma\!\left(\zeta + \mbox{
$\!\!\frac{1}{2}$}\right)\!\!~_{1\!}F_1(-\zeta, 1- 2\zeta,
A)\right].\label{A123}
\\[-2mm] \nonumber\end{eqnarray}
The integration is valid for $\Re~\zeta >0$ and $\Re~\zeta < 0$.
The result allows us to write $\Phi^{(-)}_v$ as
 \begin{eqnarray}\nonumber \\[-3mm]
\Phi^{(-)}_v ~= ~\frac{1}{2}  ~- ~h^{(-)}_1(\mu,t/\delta, A) ~- \
h^{(-)}_2(\mu,t/\delta, A)\, ,\label{phi-minus}
\end{eqnarray}
with
 \begin{eqnarray}\nonumber \\[-3mm]
h^{(-)}_1(\mu,t/\delta, A) ~&=& \
\frac{1}{2\Gamma(t/\delta)\sqrt{\pi} }~\sum_{\mathrm{Res}}
\frac{\Gamma(\zeta + t/\delta)}{\zeta}\left(8\mu\right)^{-\zeta}
 \Gamma\!\left(\mbox{
$\!\!\frac{1}{2}$} -\zeta\right) \!~_{1\!}F_1(\zeta, 1 + 2\zeta,
A)\, ,\nonumber \\[2mm]
h^{(-)}_2(\mu,t/\delta, A) ~&=& \
\frac{1}{2\Gamma(t/\delta)\sqrt{\pi} }~\sum_{\mathrm{Res}}
\frac{\Gamma(\zeta + t/\delta)}{{\zeta}}
\left(\frac{A^2\mu}{2}\right)^{-\zeta} \Gamma\!\left(\zeta + \mbox{
$\!\!\frac{1}{2}$}\right) \!~_{1\!}F_1(-\zeta, 1- 2\zeta, A)\,
.\nonumber \\
\\[-2mm] \nonumber\end{eqnarray}
Mellin's fundamental strip for $\zeta$ can be conveniently chosen in
$h^{(-)}_1$ as $0 < \zeta < 1/2$, and in $h^{(-)}_2$ as $-1/2 < \zeta < 0$. As
previously in the calculation of $\Phi^{(+)}_v$, the contour of the $s$-integration
for $h^{(-)}_2$  is closed on the left.
Thus we obtain by analogy with $\Phi^{(+)}_v$ that
the final expression has the form (\ref{A13}).

 In the small-$ \delta $ limit where
$(t/\delta)_{z} \rightarrow (t/\delta)^z$,
the third sum in (\ref{A13}) goes to zero.  In such a situation
\begin{eqnarray}\nonumber \\[-3mm]
\Phi^{(-)}_v ~= ~\frac{1}{2}\left[1 ~+ ~\sqrt{\frac{2}{\pi}}~y_{\bar v}^-
 \!~_{1\!}F_1\left(\frac{1}{2}, \frac{3}{2}, - \frac{\left(
y_{\bar v}^-\right)^2}{2}\right) \right] ~= ~\Phi(y_{\bar v}^- ) \,
,
\\[-3mm] \nonumber\end{eqnarray}
i.e., we regain the cumulative normal distribution.
In the limit $A\rightarrow 0$, one finds that $\Phi^{(-)}_v \rightarrow
\Phi( y_{\bar v}^- )|_{A=0}$.
%

\section*{Appendix~C: Information Entropy
and Tsallis Statistics}\label{SEc2}

 A useful conceptual frame that allows to generate
important classes of distributions is based on information
entropies. Information entropies generally represent measures of
uncertainty inherent in a distribution describing a given
statistical or information-theoretical system. Central role of
information entropies is in that they serve as inference functionals
whose extremalization subject to certain constraint conditions
(known as prior information), yields the MaxEnt distribution.
Importance of information entropies as tools for inductive inference
(i.e., inference where new information is given in terms of expected
values) was emphasized by many authors~\cite{Fad1}.

 Among the many possible information entropies one may
focus attention on two examples, first on R\'enyi's entropy
\begin{eqnarray}
{\mathcal{S}}_q^{(R)} ~= ~\frac{1}{1-q} \log \sum_i p_i^q\,
,\;\;\;\;\;\;\;\; q > 0\, ,\label{II1}
\end{eqnarray}
and second on the Tsallis-Havrda-Charv\'{a}t (THC) entropy
\begin{eqnarray}
{\mathcal{S}}_q^{(THC)} ~= ~\frac{1}{1-q} \left(\sum_i p_i^q
-1\right)\, ,\;\;\;\;\;\;\;\; q > 0\, .\label{II2}
\end{eqnarray}
The discrete distribution ${\mathcal{P}} = \{p_i\}$ is usually
associated with a discrete set of micro-states in statistical
physics, or set of all transmittable messages in information theory.
In the limit $q\rightarrow 1$, the two entropies coincide with each
other, both reducing to the thermodynamic Shannon-Gibbs entropy
\begin{eqnarray}
S ~= ~- \sum_{i} p_i \log p_i\, .
\end{eqnarray}
Thus the parameter $q-1$ characterizes the departure from the usual
Boltzmann-Gibbs statistics or from Shannonian information theory.

 It is well known~\cite{jaynes57} that within the context
of Shannonian information theory the laws of equilibrium statistical
mechanics can be viewed as {\em inferences} based entirely on prior
information that is given in terms of expected values of
energy and number of particles, energy and volume, energy and
angular momentum, etc.
 For the sake of simplicity we shall consider here only the
analog of canonical ensembles, where the prior information is
characterized by a fixed energy expectation value. The corresponding
MaxEnt distributions for ${\mathcal{S}}_q^{(R)}$ and
${\mathcal{S}}_q^{(THC)}$ can be obtained by extremizing the
associated inference functionals
\begin{eqnarray}
L_q^{(R)}({\mathcal{P}}) ~&=& ~\frac{1}{1-q}\log\sum_{i}p_i^q ~-
~\alpha \sum_i
p_i ~- ~\beta \langle H \rangle_r\, , \nonumber \\
L_q^{(THC)}({\mathcal{P}}) ~&=& ~\frac{1}{1-q}\left(\sum_{i}p_i^q
-1 \right)~- ~\alpha \sum_i p_i ~- ~\beta \langle H \rangle_r\,
,
\end{eqnarray}
where $\alpha$ and $\beta$ are Lagrange multipliers, the latter
being the analog of the inverse temperature in natural units.
The subscript $r$ on the energy expectation value $\langle H
\rangle$ distinguishes two conceptually different approaches. In
information theory one typically uses the linear mean, i.e.,
\begin{eqnarray}
\langle H \rangle_1 ~\equiv ~\langle H \rangle_{r=1} ~= ~\sum_i
p_i E_i\, ,
\end{eqnarray}
while in non-extensive thermostatistics it is customary to utilize a
non-linear mean
\begin{eqnarray}
\langle H \rangle_q ~\equiv ~\langle H \rangle_{r=q} ~= ~\sum_i
P_i(q) E_i,\, \;\;\;\; \mbox{with} \;\;\;\;P_i(q) ~\equiv \
\frac{p_i^q}{\sum_i p_i^q}\, , \;\;\; \sum_i p_i  ~= ~1\, .
\end{eqnarray}
The distribution $P_i(q)$ is called {\em escort\/} or {\em
zooming\/} distribution and it has its origin in chaotic
dynamics~\cite{beck} and in the physics of multifractals~\cite{PJ1}.
Simple analysis reveals~\cite{11} that
\begin{eqnarray}\nonumber \\[-4mm]
\frac{\delta L_q^{(R)}({\mathcal{P}})}{\delta p_i} ~= ~0 \
\Rightarrow \left\{
              \begin{array}{ll}
                {p}_i^{(1)} ~= ~Z_R^{-1}\left[ 1 ~- ~\tilde{\beta}
                (q-1)\Delta E_i
 \right]^{1/(q-1)}, &\;\; {\mbox{for}}
\;\; \langle H \rangle_{r=1}\, ,
\\[3mm]
                p^{(2)}_i~= ~Z_R^{-1}\left[ 1 ~- ~\beta(1-q)
                \Delta E_i \right]^{1/(1-q)} ,
        &\;\; \hbox{for} \; \; \langle H
\rangle_{r=q}\, .
              \end{array}
            \right. \label{2.7}
\\[-1mm] \nonumber
\end{eqnarray}
Here $\tilde{\beta} = \beta/q$ and $\Delta E_i = E_i - \langle H
\rangle_r$. By the same token one obtains for the THC case~\cite{11}
\begin{eqnarray}\nonumber \\[-3mm]
\frac{\delta L_q^{(THC)}({\mathcal{P}})}{\delta p_i} ~= ~0 \
\Rightarrow \left\{
              \begin{array}{ll}
{p}^{(1)}_i ~= ~Z_{THC}^{-1}\left[1 ~- ~ \tilde{\beta}^* (q-1)
\Delta E_i \right]^{1/(q-1)}, &\;\; {\mbox{for}} \;\; \langle H
\rangle_{r=1}\, ,\\[3mm]
p^{(2)}_i ~= ~Z_{THC}^{-1}\left[1 ~- ~\beta^{*}(1-q) \Delta E_i
\right]^{1/(1-q)}, &\;\; {\mbox{for}} \;\; \langle H \rangle_{r=
q}\, ,
                     \end{array}
            \right. \label{2.8}\\[-1mm] \nonumber
\end{eqnarray}
 with $\beta^* ~= ~\beta/\sum_i p_i^q$ and
$\tilde{\beta}^* ~= ~\tilde{\beta}/\sum_i p_i^q$. So in contrast
to (\ref{2.7}), the THC MaxEnt distributions are self-referential.
Generalized distributions of the form (\ref{2.7}) and (\ref{2.8})
are known as Tsallis distributions and they appear in
numerous statistical systems~\cite{Ts2a}.
For historical reasons  is ${\mathcal{P}}^{(1)} = \{ p_i^{(1)}\}$ in
(\ref{2.8}) also known as the Bashkirov's $1$-st version of
thermostatistics, while ${\mathcal{P}}^{(2)}= \{ p_i^{(2)}\}$ in
(\ref{2.8}) is called the Tsallis' $3$rd version of
thermostatistics.
An important feature of Tsallis distributions is that they are
invariant under uniform shifts $\varepsilon$ of the energy spectrum.
So one can always choose to work directly with $E_i$ rather than
$\Delta E_i$.

%


\begin{thebibliography}{99}
\bibitem{CallE} {European call options are options that give the right to purchase a
unit of stock for a strike price $E$ only at expiration date.}
\end{thebibliography}
\end{document}